\documentclass[aps,amsfonts,reprint,tightenlines,amssymb,superscriptaddress,onecolumn]{revtex4-2}  

\usepackage{graphicx}
\usepackage{dcolumn}
\usepackage{bm}
\usepackage{mathtools}

\usepackage{qcircuit}
\usepackage[caption=false]{subfig}

\usepackage{algorithm}
\usepackage[noend]{algpseudocode}
\algrenewcommand\algorithmicdo{}

\makeatletter
\renewcommand{\ALG@name}{Procedure}
\makeatother

\usepackage[linktocpage=true,
  colorlinks=true, 
  pdfborder={0 0 0},
  linkcolor=blue,
  citecolor=red,
  filecolor=yellow,
  urlcolor=blue,
  bookmarks,
  pdfauthor={},
]{hyperref}

\usepackage{ifthen}
\newcounter{is_qcircuit_used}
\begin{document}

\preprint{APS/123-QED}

\title{Acceleration of probabilistic imaginary-time evolution method combined with quantum amplitude amplification}

\author{Hirofumi Nishi}
\email{nishi.h.ac@m.titech.ac.jp}
\affiliation{
Laboratory for Materials and Structures,
Institute of Innovative Research,
Tokyo Institute of Technology,
Yokohama 226-8503,
Japan
}

\affiliation{
Quemix Inc.,
Taiyo Life Nihombashi Building,
2-11-2,
Nihombashi Chuo-ku, 
Tokyo 103-0027,
Japan
}

\author{Taichi Kosugi}
\affiliation{
Laboratory for Materials and Structures,
Institute of Innovative Research,
Tokyo Institute of Technology,
Yokohama 226-8503,
Japan
}

\affiliation{
Quemix Inc.,
Taiyo Life Nihombashi Building,
2-11-2,
Nihombashi Chuo-ku, 
Tokyo 103-0027,
Japan
}

\author{Yusuke Nishiya}
\affiliation{
Laboratory for Materials and Structures,
Institute of Innovative Research,
Tokyo Institute of Technology,
Yokohama 226-8503,
Japan
}

\affiliation{
Quemix Inc.,
Taiyo Life Nihombashi Building,
2-11-2,
Nihombashi Chuo-ku, 
Tokyo 103-0027,
Japan
}

\author{Yu-ichiro Matsushita}

\affiliation{
Laboratory for Materials and Structures,
Institute of Innovative Research,
Tokyo Institute of Technology,
Yokohama 226-8503,
Japan
}

\affiliation{
Quemix Inc.,
Taiyo Life Nihombashi Building,
2-11-2,
Nihombashi Chuo-ku, 
Tokyo 103-0027,
Japan
}

\affiliation{
Quantum Material and Applications Research Center,
National Institutes for Quantum Science and Technology,
2-12-1, Ookayama, Meguro-ku, Tokyo 152-8552, Japan
}

\date{\today}

\begin{abstract}
A probabilistic imaginary-time evolution (PITE) method was proposed as a nonvariational method to obtain a ground state on a quantum computer. In this formalism, the success probability of obtaining all imaginary-time evolution operators acting on the initial state decreases as the imaginary time proceeds. To alleviate the undesirable nature, we propose quantum circuits for PITE combined with the quantum amplitude amplification (QAA) method. We reduce the circuit depth in the combined circuit with QAA by introducing a pre-amplification operator. 
We successfully demonstrated that the combination of PITE and QAA works efficiently and reported a case in which the quantum acceleration is achieved.
Additionally, we have found that by optimizing a parameter of PITE,  we can reduce the number of QAA operations and that deterministic imaginary-time evolution (deterministic ITE) can be achieved which avoids the probabilistic nature of PITE.
We applied the deterministic ITE procedure to multiple imaginary-time steps and discussed the computational cost for the circuits. Finally, as an example, we demonstrate the numerical results of the PITE circuit combined with QAA in the first- and second-quantized Hamiltonians.
\end{abstract}

\maketitle 
\clearpage

\section{Introduction}
\label{sec:introduction}
The idea of Feynman \cite{Feynman1982} that quantum computers can efficiently simulate  quantum systems has been theoretically demonstrated in the first- and second-quantized Hamiltonians by Lloyd \cite{Lloyd1996Science, Abrams1997PRL}. Ground state calculations are particularly important in quantum simulations. It has been shown that the ground state energy can be obtained by quantum phase estimation (QPE) \cite{Kitaev1995arXiv, Abrams1999PRL}, which is implemented using real-time evolution operators. The ground-state energy is retrieved with a probability $|\langle \psi|\phi_{\mathrm{gs}}\rangle|^2$, which is the modulus squared of the overlap between the input state $|\psi\rangle$ and the ground state $|\phi_{\mathrm{gs}}\rangle$.
It is, therefore, necessary to develop an algorithm for preparing the input state as a good approximation of the ground state.
Two major directions for the state preparation are reported:
Aspuru-Guzik et al. proposed a quantum algorithm based on adiabatic time evolution \cite{Kadowaki1998PRE, Farhi2000arXiv} in reference \cite{AspuruGuzik2005Science}.
The other approach is a filtering method that decreases the components of the states that differ from the ground state \cite{Poulin2009PRL, Ge2019JMP, Lin2020Quantum, Choi2021PRL}.

The imaginary-time evolution (ITE) method is thought of as an efficient algorithm to obtain the ground state of a quantum system since the ITE method receives the benefit of the exponential decay of the high-energy states.
On the contrary, implementing the ITE operator as a gate sequence on a quantum computer is impossible due to its nonunitary.
One of the algorithms for realizing the ITE operator is to introduce a parameterized quantum circuit and to optimize the parameters contained in the quantum circuit by classical optimization procedure \cite{yuan2019theory,mcardle2019variational,jones2019variational}. This is so-called variational imaginary-time evolution (VITE). 
The VITE method is a type of variational quantum eigensolver (VQE) \cite{Peruzzo2014Ncom}. 
We obtain the ground state within the framework by varying  parameters in the ansatz using imaginary-time evolution. The parameter evolution is determined by solving a linear equation constructed from multiple measurements. 
It is possible to simulate non-equilibrium dynamics using the VITE method.
Moreover, there are several applications of the VITE method, e.g., the Gibbs partition function \cite{wu2022estimating,matsumoto2022calculation} and the Boltzman machine in a kind of machine learning \cite{zoufal2021variational,shingu2021boltzmann}.

Another ITE method on a quantum computer, called the quantum imaginary-time evolution (QITE) method, is devised to avoid the parameterized circuit \cite{motta2020determining}. In the QITE method, we construct an approximated unitary operator to reproduce the state acted on by the ITE operator in the first order in imaginary time step, $\Delta\tau$. This approximated unitary is determined by the solution of a linear equation solved by a classical computer. We have to construct the approximated unitary in every imaginary time step. The size of the quantum circuit depth grows linearly with the number of imaginary time steps. There are proceeding works to tackle the problem \cite{nishi2021implementation,gomes2020efficient}. Furthermore, there are several reports of the application of the QITE method for excited state \cite{yeter2020practical}, correlation function at finite temperature \cite{sun2021quantum}, and non-equilibrium dynamics \cite{kamakari2022digital}. 
However, these VITE and QITE require the many-times of measurement to construct the linear equation and computational power to solve the huge linear equation by a classical computer when we solve a large system. Someday the limitation of their applications will be reached by the classical computer side.

Recently, a new ITE method was proposed that avoids both parameterized and approximate circuits \cite{PITE}.
This method introduces an ancillary qubit to insert the ITE operator into an extended unitary operator. 
By using the ancillary qubit, only when we measure the ancillary qubit as the desired state, we get the state acted on by the ITE operator successfully.
Thus, this method is called the probabilistic imaginary-time evolution (PITE) method \cite{PITE}.
Actually, a probabilistic ITE implementation  on a quantum computer was proposed in early 2000, \cite{gingrich2004non,terashima2005nonunitary} and its development is in reference \cite{liu2021probabilistic,lin2021real}.
However, the previous PITE method needs to decompose the nonunitary with the use of a singular value decomposition (SVD) in advance, which is not applicable when the number of qubits is large.
In reference \cite{PITE}, one ancillary qubit is introduced to implement the ITE operator, and escaping the use of SVD. 
Reference \cite{PITE} reported specific quantum circuits for the exact and the first-order approximate ITE operator. Especially, for practical purposes, it is desirable to use the approximate circuit in the first-order for small imaginary-time $\Delta \tau$, because it is composed of forward and backward real-time evolution and we can use any real-time evolution algorithm for those parts. 
The PITE method is applied to electronic structure calculations and structure optimization problems in the field of quantum chemistry based on the first-quantized formula \cite{PITE,Kosugi2022arXiv}.

A disadvantage of PITE is attributed to its probabilistic nature, which reduces the success probability of obtaining the state acted on by the ITE operator throughout imaginary-time steps. Recall that only when we measure the ancillary qubit as the desired state, we successfully realize the ITE operation.
In this work, we cope with this problem by using the quantum amplitude amplification (QAA) method. 
In QAA, by repeatedly operating the amplitude amplification operator to an effective two-dimensional space spanned by the desired state and the orthogonal states, we can amplify the amplitude of the desired state. 
This paper shows efficient implementation of quantum circuits for PITE combined with QAA. 
Moreover, we found that PITE scheme can be modified to an even non-probabilistic or deterministic ITE method, by making full use of a separable state between an ancillary and working qubits after the QAA operation. 

The structure of this paper is given as follows. 
It begins with a section \ref{sec:method}, with a brief explanation of the PITE method and the QAA technique. 
In section\ref{sec:circuit}, we provide quantum circuits for PITE combined with QAA, which is the main finding.
From the developed quantum circuit, we present the numerical results in section \ref{sec:results}.
Section \ref{sec:conclusions} is devoted to the conclusion of this manuscript.

\section{Theory}
\label{sec:method}
\subsection{PITE}
\label{sec:method_pite}
\subsubsection{Nonunitary}
We consider the implementation of a nonunitary Hermitian operator $\mathcal{M}$ on an $n$-qubit system. 
We will focus mainly on an ITE operator as the nonunitary $\mathcal{M}$, but discuss a more general case in this paragraph. 
To implement the nonunitary on a quantum computer, we extend the $n$-qubit system to an $(n+1)$-qubit system by introducing an ancillary qubit as
\begin{gather}
    \mathcal{U}_{\mathcal{M}}
    \equiv
    \begin{pmatrix}
        \mathcal{M} & \sqrt{1-\mathcal{M}^2} \\
        \sqrt{1-\mathcal{M}^2} & -\mathcal{M} 
    \end{pmatrix}_a ,
\end{gather}
where the matrix labeled by a subscript "$a$" designates its action depending on the state of the ancillary qubit.
The approach of embedding nonunitary as a submatrix of a unitary is known as block encoding \cite{Martyn2021PRXQuantum}.
Let the initial state of the $n$-qubit system be $|\psi\rangle$, from which we get a quantum state depending on the ancillary state from the unitary as, 
\begin{gather}
\begin{aligned}
    \mathcal{U}_{\mathcal{M}}
    |\psi\rangle \otimes|0\rangle 
    &=
    \mathcal{M}|\psi\rangle \otimes|0\rangle
    +
    \sqrt{1-\mathcal{M}^{2}}|\psi\rangle \otimes|1\rangle 
    \\
    \mathcal{U}_{\mathcal{M}}
    |\psi\rangle \otimes|1\rangle 
    &=
    -\mathcal{M}|\psi\rangle \otimes|1\rangle
    +
    \sqrt{1-\mathcal{M}^{2}}|\psi\rangle \otimes|0\rangle . 
\end{aligned}
\end{gather}
By post-selecting the ancillary qubit being $|0\rangle$ state, we can obtain the state $\mathcal{M}|\psi\rangle$ acted on by the nonunitary operator with a probability $\mathbb{P}_0=\langle\psi|\mathcal{M}^{2}|\psi\rangle$.
When we observe the ancillary $|1\rangle$ state,
which we refer to as the failure state,
the state of the working qubits has become $\sqrt{1-\mathcal{M}^2}|\psi\rangle$.

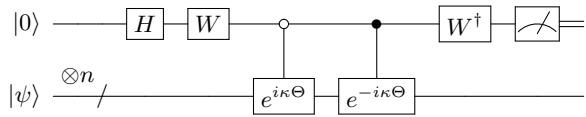
\begin{figure}
\centering
\mbox{ 
\Qcircuit @C=1em @R=1.5em { 
\lstick{ \ | 0 \rangle}  & \qw                    & \qw & \gate{H} & \gate{W} & \ctrlo{1}           & \ctrl{1} &  \gate{W^\dagger} & \meter & \cw \\
\lstick{ \ | \psi \rangle} & \ustick{\otimes n} \qw & {/} \qw  & \qw & \qw      & \gate{e^{i\kappa\Theta}} & \gate{e^{-i\kappa\Theta}} & \qw & \qw & \qw \\
}
} 
\caption{
Quantum circuit $\mathcal{C}_{\mathcal{M}}$ for probabilistic operation of nonunitary $\mathcal{M}$ that acts on an input $n$-qubit state $|\psi\rangle$. $H$ is the Hadamard gate.
}
\label{circuit:imag_evol_as_part_of_real_evol}
\end{figure}

This embedded unitary $\mathcal{U}_{\mathcal{M}}$ can be decomposed as
\begin{gather}
    \mathcal{U}_{\mathcal{M}}
    =
    ( I_{2^n} \otimes W^{\dagger} )
    \begin{pmatrix}
        e^{i \kappa \Theta} & 0 \\
        0 & e^{-i \kappa \Theta}
    \end{pmatrix}_a
    ( I_{2^n}\otimes WH ) ,
\end{gather}
where $I_{2^n}$ is the identity matrix for an $n$-qubit system and we introduced a Hermitian operator $\Theta$ for an $n$-qubit system defined by the nonunitary $\mathcal{M}$ as 
\begin{gather}
    \Theta 
    \equiv 
    \arccos \frac{\mathcal{M}+\sqrt{1-\mathcal{M}^{2}}}{\sqrt{2}}.
\end{gather}
We see the inequality $\sin \Theta = (\mathcal{M}-\sqrt{1-\mathcal{M}^2})/\sqrt{2} \geq 0$ derived from the range of principal value of the arccosine function as $0 \leq \Theta \leq \pi$. 
To ensure the inequality, we define $\kappa = \operatorname{sgn}(\|\mathcal{M}\| - 1/\sqrt{2})$, which leads to 
$\cos \Theta = (\mathcal{M}+\sqrt{1-\mathcal{M}^2})/\sqrt{2}$
and 
$\sin \kappa \Theta = (\mathcal{M}-\sqrt{1-\mathcal{M}^2})/\sqrt{2} $ .
Thus, by using the controlled unitaries $e^{\pm i \kappa \Theta}$ and the single-qubit gates,
\begin{gather}
    W 
    \equiv 
    \frac{1}{\sqrt{2}}
    \left(\begin{array}{cc}
        1 & -i \\
        1 & i
    \end{array}\right) ,
\end{gather}
the quantum circuit $\mathcal{C}_{\mathcal{M}}$ for the unitary $\mathcal{U}_{\mathcal{M}}$ can be constructed as shown in Fig. \ref{circuit:imag_evol_as_part_of_real_evol}.

\subsubsection{Approximate PITE}
\begin{figure}
\centering

\subfloat[]{
\Qcircuit @C=1em @R=1.5em { 
\lstick{\ | 0 \rangle}  & \qw                    & \qw & \gate{H} & \gate{W} & \ctrlo{1}           & \ctrl{1} & \gate{R_z (-2 \theta)} & \gate{W^\dagger} & \meter & \cw \\
\lstick{\ | \psi \rangle} & \ustick{\otimes n} \qw & {/} \qw  & \qw & \qw      & \gate{U_{\mathrm{RTE}}} & \gate{U_{\mathrm{RTE}}^\dagger} & \qw & \qw & \qw & \qw \\
} 
}

\subfloat[]{
\Qcircuit @C=1em @R=1.5em { 
\lstick{\ | 0 \rangle}  & \qw                    & \qw      & \gate{H} & \gate{W}                & \ctrl{1} & \gate{R_z (-2 \theta)} & \gate{W^\dagger} & \meter & \cw \\
\lstick{\ | \psi \rangle} & \ustick{\otimes n} \qw & {/} \qw  & \qw      & \gate{U_{\mathrm{RTE}}} & \gate{U_{\mathrm{RTE}}^{\dagger 2}} & \qw & \qw & \qw & \qw \\
} 
}

\caption{
(a) Quantum circuit $\mathcal{C}_{\mathrm{PITE}}^{(1)}$ equivalent to $\mathcal{C}_{\mathrm{PITE}}$ in the first order in $\Delta \tau$.
$U_{\mathrm{RTE}} \equiv U_{\mathrm{RTE}} (s_1 \Delta \tau)$ is used in this figure.
(b) The equivalent circuit to (a).
}
\label{circuit:imag_evol_as_part_of_real_evol:evolution_1st_order}
\end{figure}
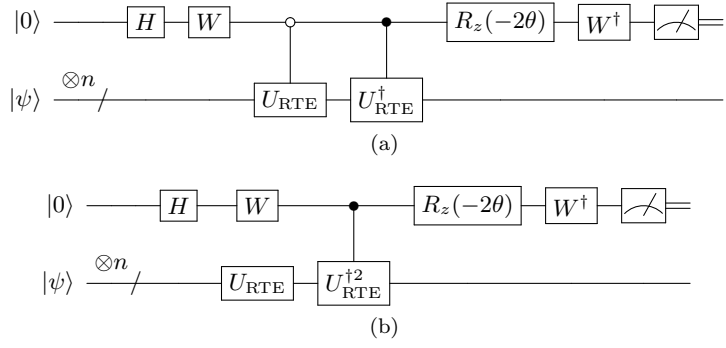

We want to decompose the unitary operator $e^{\pm i \kappa\Theta}$ into basic gates for implementing the nonunitary $\mathcal{M}$. 
However, since it is generally challenging to decompose this unitary into basic gates, we introduce an approximation to the unitary.
So far we have discussed the general cases for the nonunitary operator $\mathcal{M}$. Hereafter let us discuss a specific case for $\mathcal{M}$ being the ITE operator.
For the $n$-qubit system with a Hamiltonian $\mathcal{H}$, the ITE operator for a small imaginary-time step $\Delta\tau$ is given as $e^{-\mathcal{H} \Delta \tau}$. 
We consider a nonunitary Hermitian operator as 
\begin{gather}
    \mathcal{M} 
    =
    \gamma e^{-\mathcal{H} \Delta \tau} ,
\end{gather}
with a real constant $\gamma$ satisfying $0<\gamma<1, \gamma\neq 1/\sqrt{2}$.
These conditions allow us to conduct Taylor expansion with avoiding singularity.
Accordingly, the Hermitian operator $\Theta$ is expanded as
\begin{gather}
    \kappa \Theta
    =
    \theta -\mathcal{H} s \Delta \tau
    +
    \mathcal{O}\left(\Delta \tau^{2}\right),
    \label{PITEwithQAA:Theta_taylor_1st_order}
\end{gather}
with the coefficients 
$
    \theta
    \equiv 
    \kappa \arccos \left[ (\gamma+\sqrt{1-\gamma^{2}})/\sqrt{2} \right]
$
and
\begin{gather}
    s \equiv \frac{\gamma}{\sqrt{1-\gamma^{2}}}.
\label{PITE:definition_of_s}
\end{gather}
Due to the transformation of the arccosine function $\Theta$ into another function $\Theta^{(1)}$ by Taylor expansion, the range of principal value for $\Theta^{(1)}$ changes from $0 \leq \Theta \leq \pi$ to any real value.
In addition, we take $\kappa = \operatorname{sgn}(\gamma-1/\sqrt{2})$, where eigenvalues of the Hamiltonian are shifted to be positive. 
This shift is justified by introducing new parameter $\gamma^{\prime}$ as $\gamma = \gamma^{\prime} e^{-E_{0}\Delta\tau}$, where $\gamma^{\prime}$ satisfies the condition $0<\gamma^{\prime}<1, \gamma^{\prime}\neq 1/\sqrt{2}$ and $E_0$ is a real positive constant. 
The quantum circuit $\mathcal{C}_{\mathrm{PITE}}^{(1)}$ of the approximate PITE operator $\mathcal{U}_{\mathrm{PITE}}^{(1)}$ is shown in Fig. \ref{circuit:imag_evol_as_part_of_real_evol:evolution_1st_order}.
We denote the quantum circuit for the exact PITE by $\mathcal{U}_{\mathcal{M}=\gamma e^{-\Delta \tau \mathcal{H}}} = \mathcal{U}_{\mathrm{PITE}}$.
We clearly see that the approximate PITE circuit can be easily implemented by using forward and backward real-time evolution operators and single-qubit gates.
The output state through the quantum circuit in Fig.  \ref{circuit:imag_evol_as_part_of_real_evol:evolution_1st_order} is derived as
\begin{gather}
    \gamma(1-\mathcal{H} \Delta \tau)|\psi\rangle \otimes|0\rangle 
    +
    \left(
        \sqrt{1-\gamma^{2}}+\frac{\gamma^{2}}{\sqrt{1-\gamma^{2}}} \mathcal{H} \Delta \tau
    \right)
    |\psi\rangle \otimes|1\rangle
    +
    \mathcal{O}\left(\Delta \tau^{2}\right).
\label{PITEwithQAA:output_state_1st_order}
\end{gather}
After obtaining the ground state via imaginary-time evolution, the energy eigenvalue of the Hamiltonian $\mathcal{H}$ can be obtained by performing QPE \cite{Abrams1999PRL}. 
In addition, the one-body Green's function \cite{kosugi2020construction} and linear-response function \cite{kosugi2020linear} can also be obtained in the case of the second-quantized Hamiltonian.

\subsection{PITE combined with QAA}
\label{sec:method_pitewqaa}
The QAA method has been proposed in reference \cite{bib:4884, bib:4878}. QAA is a generalization of Grover's search algorithm \cite{grover1996fast,grover1997quantum}. 
Recently, further generalization of QAA has been studied, which employs polynomial transformation called quantum signal processing (QSP) \cite{Low2017PRL} such that the operation of the amplitude amplification can be thought of as a rotation on the Bloch sphere \cite{Gilyen2019ACM, Martyn2021PRXQuantum}.   

Let an $n$-qubit input state be prepared as $| \psi \rangle = U_{\mathrm{ref}} | 0 \rangle^{\otimes n}$, where $U_{\mathrm{ref}}$ is called a reference circuit.
The state of the $(n+1)$-qubit system, consisting of the $n$-qubit register and one ancillary qubit, before a measurement is written as
\begin{align}
    | \Psi \rangle
    =
        \mathcal{U}_{\mathrm{PITE}}
        (| \psi \rangle \otimes | 0 \rangle)
    =
        a
        \underbrace{
            | \psi_{\mathrm{success}} \rangle \otimes | 0 \rangle
        }_{\equiv | \Psi_{\mathrm{good}} \rangle}
        +
        \sqrt{1-a^2}
        \underbrace{
            | \psi_{\mathrm{failure}} \rangle \otimes | 1 \rangle
        }_{\equiv | \Psi_{\mathrm{bad}} \rangle}
        ,
\end{align}
where $a$ is real and the states $| \Psi_{\mathrm{good}} \rangle$ and $| \psi_{\mathrm{success}} \rangle$ are normalized as
$
\langle \Psi_{\mathrm{good}} | \Psi_{\mathrm{good}} \rangle
=
\langle \psi_{\mathrm{success}} | \psi_{\mathrm{success}} \rangle
= 1.
$
Since the success state, denoted by $| \psi_{\mathrm{success}} \rangle$, is entangled with the ancillary qubit state $| 0 \rangle$, we define the good state $|\Psi_{\mathrm{good}} \rangle$ and the bad state $|\Psi_{\mathrm{bad}} \rangle$ depending on the ancillary qubit as $| 0 \rangle$ or $| 1 \rangle$, respectively.
The unitary acting on the $|0\rangle^{\otimes n} \otimes | 0 \rangle$ state to generate $|\Psi\rangle$ is defined as
\begin{gather}
    \mathcal{U}_{\mathrm{PITE+ref}}
    \equiv
    \mathcal{U}_{\mathrm{PITE}}
    (U_{\mathrm{ref}} \otimes I_2) .
\end{gather}

We introduce a unitary operator $
    S_\chi(\phi)
    =
    e^{i\phi|\Psi_{\mathrm{good}}\rangle\langle\Psi_{\mathrm{good}}|}
$, named an oracle, that causes a phase shift only to the good-state component.
Another unitary $S_{0}(\phi)=e^{i\phi|0\rangle\langle 0|}$ named zero reflection changes the phase only of $| 0 \rangle^{\otimes (n + 1)}.$
Using these two unitaries, the oracle $S_\chi$ and the zero reflection $S_0$, we define an amplification operator
in the $i$th step as
\begin{gather}
    Q(\phi_{2i-1}, \phi_{2i}) 
    \equiv
        -
        \mathcal{U}_{\mathrm{PITE+ref}}
        S_0(\phi_{2i-1})
        \mathcal{U}_{\mathrm{PITE+ref}}^\dagger
        S_{\chi}(\phi_{2i})
    \label{QITE_with_QAA:def_amplification_opr},
\end{gather}
where $\phi_{2i-1}$ and $\phi_{2i}$ are the rotational angles for the zero reflection and the oracle,
respectively.
By defining $\theta_a$ via $a=\sin \theta_a$ ($0 \leq \theta_a \leq \pi/2$),
we obtain the target state $|\Psi_{\mathrm{good}}\rangle$ after $m$ times operation of the amplification operator with the probability,
\begin{gather}
    \langle \Psi_{\mathrm{good}} |
    \left[ \prod_{i=1}^{m} Q(\phi_{2i-1}, \phi_{2i}) \right]
    |\Psi \rangle
    =
    \mathrm{Poly} (a)
    ,
\label{eq:qaa_qsp_success_probability}
\end{gather}
where $\mathrm{Poly}(a)$ is a polynomial of $a$ so that 
 the degree of the polynomial is at most $2m$ and $\mathrm{Poly}(a)$ has odd parity. 
Also, the sum of the probabilities for getting the state $|\Psi_{\mathrm{good}}\rangle$ and $|\Psi_{\mathrm{bad}}\rangle$ should be one (conservation of probability).
The original QAA is the case of $\phi_i = \pm \pi$. 
In this case, the initial state $|\Psi\rangle$ changes via QAA as,
\begin{gather}
    Q^{m} (\pi,\pi)
    | \Psi \rangle
    =
        \sin [ (2 m + 1) \theta_a]
        | \Psi_{\mathrm{good}} \rangle
        +
        \cos [ (2 m + 1) \theta_a]
        | \Psi_{\mathrm{bad}} \rangle .
    \label{eq:qaa_quantum state}
\end{gather}
The weight of the good state in the amplified state is equal to $a_m \equiv \sin^2 [ (2 m + 1) \theta_a]$ (from the definition, $a_{m=0}=a^2$).
Thus, the probability goes to the maximum and minimum  alternately in increasing the number of repetitions $m$ of QAA. 
When we choose the number of operations $m$ to maximize $a_m$, the probability to observe the good state is maximized.
If we regard 
the number $m^{*}$ of operations that maximizes the weight of the good state to take continuous values,
it is given as
\begin{gather}
	m^{*} = 
    \left\lfloor \frac{(2n+1)\pi}{4 \sin^{-1} a} \right\rfloor  ,
    \label{PITEwithQAA:optimal_repetition_of_QAA}
\end{gather}
where $n$ is an integer.
It is necessary to estimate $a$ to obtain the optimal number of the repetition $m^{*}$ for QAA by using quantum amplitude estimation (QAE) \cite{bib:4878,bib:5145}.

As shown in Eq. (\ref{eq:qaa_quantum state}), an inappropriate number of repetitions of QAA brings us even a lower success probability, the so-called "overcook."
In QAA based on QSP, by choosing the parameters $\{\phi_i\}$ properly, we can design a preferable polynomial function such that the high success probability remains even in excessive repetition of QAA. 
In such case, we can avoid estimating the weight $a_0$ of the good state.
Another way for QAA is the fixed point search that accomplishes to get the desired state with an error $1-\delta^2$ where $\delta$ is precision \cite{Yoder2014PRL}.
The QSP technique obtains the approximated sign function in Ref. \cite{Martyn2021PRXQuantum}.

\section{Methods}
\label{sec:circuit}
\subsection{First step of PITE with QAA}
\label{sec:circuit_one}
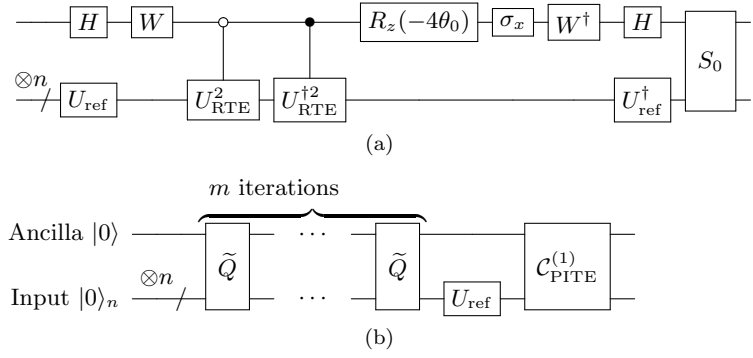
\begin{figure}
\centering

\subfloat[]{
\Qcircuit @C=.6em @R=1.5em { 
\lstick{} & \qw                    & \qw      & \gate{H}                  & \gate{W} & \ctrlo{1}                 & \ctrl{1}                            & \gate{R_z (-4 \theta_0)} & \gate{\sigma_x} & \gate{W^\dagger} & \gate{H}
& \multigate{1}{S_0} & \qw \\
\lstick{} & \ustick{\otimes n} \qw & {/} \qw  & \gate{U_{\mathrm{ref}}} & \qw      & \gate{U_{\mathrm{RTE}}^2} & \gate{U_{\mathrm{RTE}}^{\dagger 2}} & \qw                      & \qw             & \qw              & \gate{U_{\mathrm{ref}}^\dagger}  & \ghost{S_0}  & \qw \\
}
}

\subfloat[]{
\Qcircuit @C=1em @R=1.5em {
& & & & \mbox{$m$ iterations} \\
\lstick{\mathrm{Ancilla} \ | 0 \rangle}  &  \qw   & \qw & \multigate{1}{\widetilde{Q}} & \qw & \push{\cdots} & & \multigate{1}{\widetilde{Q}} & \qw & \multigate{1}{\mathcal{C}_{\mathrm{PITE}}^{(1)}} & \qw \\
\lstick{\mathrm{Input} \ | 0 \rangle_n} & \ustick{\otimes n} \qw & {/} \qw & \ghost{\widetilde{Q}}        & \qw & \push{\cdots} & & \ghost{\widetilde{Q}} & \gate{U_{\mathrm{ref}}} & \ghost{\mathcal{C}_{\mathrm{PITE}}^{(1)}} & \qw
\gategroup{2}{4}{3}{8}{.5em}{^\}}
} 
}

\caption{
(a) Quantum circuit for pre-amplification operator $\widetilde{Q}$ when we use the approximate PITE $\mathcal{U}_{\mathrm{PITE}}^{(1)}$.
(b) Overall circuit for the first step of PITE combined with QAA with the use of the quantum circuit in (a).
}
\label{circuit:PITE_with_QAA_using_preposed_ampl}
\end{figure}

As we see in Eq. (\ref{QITE_with_QAA:def_amplification_opr}), the amplification operator $Q$ includes $\mathcal{U}_{\mathrm{PITE+ref}}$ and its Hermitian conjugate.
Thus, when we directly implement the amplification operator $Q$, its circuit depth becomes more than twice the depth of $\mathcal{U}_{\mathrm{PITE+ref}}$.
In this subsection, we show that it is possible to execute QAA by a shallower circuit.  
In the case of PITE, the oracle is given as
\begin{gather}
    S_\chi(\phi)
    =
    I_{2^n} \otimes (\sigma_x Z_{\phi} \sigma_x),
    \label{QITE_with_QAA:S_chi}
\end{gather}
where $Z_{\phi} = \mathrm{diag}(1, e^{i\phi})$ is the phase shift gate.
We see that the oracle $S_\chi$ can be implemented as a single-qubit gate acting on the ancillary qubit in
Fig.~\ref{circuit:imag_evol_as_part_of_real_evol:evolution_1st_order}.
Let us consider QAA after the first step of the PITE circuit, which corresponds to substitution of $\mathcal{U}_{\mathrm{PITE+ref}}| 0 \rangle^{\otimes n+1} $ into $|\Psi \rangle$ in Eq.~(\ref{eq:qaa_qsp_success_probability}). We rewrite the unitary operation for QAA in Eq. (\ref{eq:qaa_qsp_success_probability}) as follows,
\begin{gather}
    \left[ 
        \prod_{i=1}^{m} Q(\phi_{2i-1},\phi_{2i})
    \right]
    \mathcal{U}_{\mathrm{PITE+ref}}
    =
    \left[
        \prod_{i=1}^{m}
        \left(
            -
            \mathcal{U}_{\mathrm{PITE+ref}}
            S_0(\phi_{2i-1})
            \mathcal{U}_{\mathrm{PITE+ref}}^\dagger
            S_\chi(\phi_{2i})
        \right)
    \right]
        \mathcal{U}_{\mathrm{PITE+ref}}
    \nonumber \\
    =
        \mathcal{U}_{\mathrm{PITE+ref}}
        \prod_{i=1}^{m}
        \left(
            -
            S_0(\phi_{2i-1})
            \mathcal{U}_{\mathrm{PITE+ref}}^\dagger
            S_\chi(\phi_{2i})
            \mathcal{U}_{\mathrm{PITE+ref}}
        \right)
    = 
    \mathcal{U}_{\mathrm{PITE+ref}}
    \prod_{i=1}^{m}
    \widetilde{Q}(\phi_{2i-1},\phi_{2i}),
    \label{QITE_with_QAA:Q_m_U_PITE}
\end{gather}
where we defined the pre-amplification operator
\begin{gather}
    \widetilde{Q}(\phi_{2i-1},\phi_{2i})
    \equiv
        S_0(\phi_{2i-1})
        (U_{\mathrm{ref}}^\dagger \otimes I_2)
        \mathcal{D}(\phi_{2i})
        (U_{\mathrm{ref}} \otimes I_2),
    \label{QITE_with_QAA:def_preposed_ampl_opr}
\end{gather}
and 
$
\mathcal{D}(\phi)
\equiv
-
\mathcal{U}_{\mathrm{PITE}}^{\dagger}
\left( I_{2^n} \otimes (\sigma_x Z_{\phi} \sigma_x) \right)
\mathcal{U}_{\mathrm{PITE}}.
$
In a special case of $\phi=\pm \pi$, the unitary $\mathcal{D} \equiv \mathcal{D}(\pi)$ is summarized as
\begin{gather}
    \mathcal{D}
    =
        \left( I_{2^n} \otimes (H W^\dagger\sigma_x) \right)
        \left(
            \left(e^{i\kappa\Theta}\right)^2
            \otimes
            | 0 \rangle \langle 0 |
            +
            \left(e^{-i\kappa\Theta}\right)^2
            \otimes
            | 1 \rangle \langle 1 |
        \right)
        \left( I_{2^n} \otimes (W H) \right) .
\end{gather}
When we use the approximate circuit $\mathcal{C}_{\mathrm{PITE}}^{(1)}$ in Fig. \ref{circuit:imag_evol_as_part_of_real_evol:evolution_1st_order} for practical purposes, we get
\begin{gather}
    \mathcal{D}^{(1)}
    =
        -
        \mathcal{U}_{\mathrm{PITE}}^{(1) \dagger}
        \left( I_{2^n} \otimes (\sigma_x \sigma_z \sigma_x) \right)
        \mathcal{U}_{\mathrm{PITE}}^{(1)}
    \nonumber \\
    =
        \left(
            I_{2^n} \otimes (H W^\dagger \sigma_x R_z (-4 \theta_0))
        \right)
        \Big(
            U_{\mathrm{RTE}}^2
            \otimes
            | 0 \rangle \langle 0 |
            +
            U_{\mathrm{RTE}}^{\dagger 2}
            \otimes
            | 1 \rangle \langle 1 |
        \Big)
        \left( I_{2^n} \otimes (W H) \right),
    \label{QITE_with_QAA:simplified_D_opr}
\end{gather}
where we denote $U_{\mathrm{RTE}} \equiv U_{\mathrm{RTE}} (s_1 \Delta \tau)$ for simplicity and use the relation $Z_{\phi=\pi}=\sigma_z$.
The quantum circuit for implementing the pre-amplitude amplification $\widetilde{Q}$ is shown in Fig. \ref{circuit:PITE_with_QAA_using_preposed_ampl}(a). 
Also, the whole process of PITE combined with QAA at the first imaginary-time step is depicted in Fig. \ref{circuit:PITE_with_QAA_using_preposed_ampl}(b).

%

\subsubsection{Reduction of computational cost}
\label{section:QITE_with_QAA_operation_cost}
We evaluate the computational cost for the amplification operator and the pre-amplification operator.
From Eq. (\ref{QITE_with_QAA:def_amplification_opr}) and (\ref{QITE_with_QAA:S_chi}), 
the number of gate operations of the quantum circuit for the amplification operator $Q$ is calculated as
\begin{align}
    c_Q
    &=
        2
        c_{\mathcal{U}_{\mathrm{PITE+ref}}}
        +
        c_{S_0}
        +
        c_{S_{\chi}}
    \nonumber \\
    &\approx
        2
        ( c_{\mathcal{U}_{\mathrm{PITE}}^{(1)}} + c_{U_\mathrm{ref}} )
        +
        c_{S_0},
    \label{QITE_with_QAA:cost_of_amplification_opr}
\end{align}
where $c_{U}$ is the number of operations for a unitary $U$.
By using a relation $U_{\mathrm{RTE}}^2 = U_{\mathrm{RTE}} (2 s_1 \Delta \tau)$, we can implement $U_{\mathrm{RTE}}^2$ such that the computational cost for $U_{\mathrm{RTE}}^2$ and $U_{\mathrm{RTE}}$ are equal.
The equivalence allows us to implement the pre-amplification $\widetilde{Q}$ with the computational cost,  
\begin{align}
    c_{\widetilde{Q}}
    &=
        c_{S_0}
        +
        2
        c_{U_\mathrm{ref}}
        +
        c_{\mathcal{D}}
    \nonumber \\
    &\approx
        c_{S_0}
        +
        2
        c_{U_\mathrm{ref}}
        + 
        c_{\mathcal{U}_{\mathrm{PITE}}} .
    \label{QITE_with_QAA:cost_of_preposed_ampl_opr}
\end{align}
Accordingly, when we use the pre-amplification operator $\widetilde{Q}$ for the implementation of QAA, we can reduce the computational cost by  
$m (c_{\widetilde{Q}} - c_Q) \approx m c_{\mathcal{U}_{\mathrm{PITE}}}$ 
from the circuit with the amplification operator $Q$.

\subsection{Deterministic imaginary-time evolution}
\label{sec:circuit_dite}
\subsubsection{Optimizing parameter}
As shown in the subsection \ref{sec:circuit_one}, the success probability is amplified by QAA as $a_m=\sin^2[(2m+1)\theta_a]$. 
It is necessary to estimate $\theta_a$ to maximize the success probability by using techniques such as QAE, which affects the optimal number of repetitions $m^{*}$. 
Here, the success probability for PITE without QAA is expressed as $\sin^{2}\theta_a = \gamma^{2} \langle \psi|e^{-2\mathcal{H}\Delta\tau} |\psi\rangle $.
These facts tell us that two parameters specify the success probability by QAA: the number of repetitions $m$ and the parameter $\gamma$ appearing in PITE. 
Since the latter parameter $\gamma$ has been introduced to perform Taylor expansion safely, we can choose its value freely as long as $0<\gamma<1$ and $\gamma\neq 1/\sqrt{2}$ are satisfied.
In Eq. (\ref{PITEwithQAA:optimal_repetition_of_QAA}), we derived the number of repetitions $m^*$ to maximize the success probability, but here, we instead derive the parameter $\gamma^{*}$ in the PITE operator to maximize the success probability.
The parameter $\gamma^{*}$ is calculated as
\begin{gather}
	\gamma^{*}
    =
    \frac{1}{\alpha} 
    \sin\left(
        \frac{(2n+1)\pi}{4m^{*}+2}
    \right) ,
\label{PITEwithQAA:optimal_gamma}
\end{gather}
where $\alpha$ is a normalization constant expressed as 
$
	\alpha^2 
    \equiv 
    \langle \psi|e^{-2\mathcal{H}\Delta\tau} |\psi\rangle
$.
From Eq. (\ref{PITEwithQAA:optimal_gamma}), we determine $\gamma^{*}$ satisfying the conditions  $0<\gamma^{*}<1$, $\gamma^{*}\neq 1/\sqrt{2}$ to minimize the number of repetition $m^{*}$ for QAA.
A similar idea to enhance the success probability as one by using inherent parameters in their algorithm has been presented in Ref.~\cite{liu2021probabilistic}. 
Although the fixed point search or oblivious amplification is proposed to avoid the "overcooks," the methods need multiple repetitions for designing the desired polynomial function. 
Contrary, in the present approach, we determine the optimal parameter to reduce the number of repetitions for QAA. 
Thus, the present technique has an advantage for computational costs. 

Suppose we use the fixed point search or QSP-based QAA. A larger value for parameter $\gamma$ is preferable since higher success probability before amplification leads to a rapid increase of sign function designed by QSP \cite{Martyn2021PRXQuantum, Yoder2014PRL}. 

\subsubsection{Number of QAA needed for the worst situation: A case study}
In this paragraph, we discuss how many repetition $m$ of QAA is needed focusing on a specific case of the infinitesimal success probability. From the discussion, we can deduce an interesting fact of quantum acceleration for a specific case.
When the normalization constant $\alpha$ is small enough, we take an approximation as
$\sin^{-1} a \approx a = \gamma \alpha$
in Eq. (\ref{PITEwithQAA:optimal_repetition_of_QAA}). 
Thus, we recognize the scaling of the number of repetition $m=\mathcal{O}(1/a)$, which is an already known fact in QAA \cite{bib:4878}. 
Further, for an $n$-qubit system $|\psi\rangle = \sum_{k}c_{k}|k\rangle$, the normalization constant $\alpha^2$ is expressed as
\begin{gather}
    \alpha^{2}
    =
    \langle\psi|e^{-2\Delta\tau\mathcal{H}}|\psi\rangle
    =
    \sum_{k=1}^{N} |c_k|^2 e^{-2 \lambda_{k} \Delta\tau},
\end{gather}
where $\lambda_k$ is the eigenvalue for the eigenvector $|k\rangle$ and $N=2^n$. 
Here, we consider an initial state as a homogeneous state, $|\psi\rangle = \frac{1}{\sqrt{N}}\sum_{k=1}^{N}|k\rangle$. 
In this situation, the normalization constant $\alpha^2$ becomes
\begin{gather}
    \alpha^{2}
    =
    \frac{1}{N}
    \sum_{k=1}^{N}
    e^{-2 \lambda_{k} \Delta\tau}.
\end{gather}
Let us consider a simple case that an electron is confined in a harmonic potential to pursue the number of repetition of QAA. The eigenvalues of the harmonic potential are discretized with an equal interval as $\lambda_k=\hbar \omega \left(k+\frac{1}{2}\right)$. 
The normalization constant is written as
\begin{gather}
    \alpha^2
    =
    \frac{1}{N} \sum_{k=0}^{N-1}
    e^{-2\Delta\tau \hbar \omega\left(k+\frac{1}{2}\right)}
    =
    \frac{1-e^{-2\Delta\tau \hbar \omega (N-1)}}{2N\sinh{(\Delta\tau \hbar \omega)}} 
    .
\end{gather}
From this equation, we can see that the normalization constant exponentially decays as the number of qubits increases in the homogenous input state and the harmonic potential. 
Thus, the number of repetition $m^{*}$ needed is approximated under the infinitesimal success probability $a$ from Eq. (\ref{PITEwithQAA:optimal_repetition_of_QAA}) as
\begin{gather}
    m^{*}
    \approx
    \frac{(2n+1)\pi}{4\gamma}
    \sqrt{
        \frac{2N\sinh{(\Delta\tau \hbar \omega)}}{1-e^{-2\Delta\tau \hbar \omega (N-1)}}
    }
    -
    \frac{1}{2}.
\end{gather}
We find $m^{*}=\mathcal{O}(\sqrt{N})$ even for $\max \gamma =1$ and the quadratic quantum acceleration as the same as Grover's algorithm.
Although this is just a case study, we have demonstrated that the combination of PITE and QAA achieves the quantum acceleration even in the worst situation
where we assume the limit of large number of qubits and that of infinitesimal success probability, which corresponds to the case that we do not know the correct answer at all and use the uniformly distributed among the possible answers. Of course, this is the ignorant limit for the correct answer. If we know somehow the information of the correct answer such as an approximate answer, the computational cost should be improved.

\subsubsection{PITE action on a separable state}
According to the method for optimizing the parameters in Sec. \ref{sec:circuit_dite},
PITE combined with QAA using the optimal parameter $\gamma$ brings us $|\Psi_{\mathrm{good}}\rangle$ with a 100\% probability, which in turn means a deterministic manner of the imaginary-time evolution method. Note that the deterministic PITE method makes the entanglement between working qubits and an ancillary qubit a separable state. That is, we derive that the initial state changes as
\begin{gather}
	\mathcal{U}_{\mathrm{PITE+ref}} \widetilde{Q}^{m^{*}} 
    \left(
    	|0\rangle^{\otimes n}\otimes |0\rangle
    \right)
	=
	|\psi_{\mathrm{success}}\rangle \otimes |0\rangle.
\label{eq:separable_state_by_pite}
\end{gather}
If the output state is a separable state as in the above equation, we may operate the next PITE operator to this output state without measurement of the ancillary qubit. We consider a unitary operator at this time as
\begin{gather}
    \mathcal{U}_{\mathrm{PITE}}(\gamma_2,\Delta\tau_2)
    \mathcal{U}_{\mathrm{PITE+ref}} \widetilde{Q}^{m^{*}}
    =
    \mathcal{U}_{\mathrm{PITE}}(\gamma_2,\Delta\tau_2) 
    \mathcal{U}_{\mathrm{PITE}}(\gamma^{*}_{1}, \Delta\tau_1)
    (U_{\mathrm{ref}} \otimes I_2)
    \widetilde{Q}^{m^{*}} .
\end{gather}
Note that, generally, the parameters, such as $\gamma$ and $\Delta\tau$, in each PITE step can be different from those in the other steps. Therefore, we denote $\gamma$ and $\Delta\tau$ in the $k$th step by $\gamma_k$ and $\Delta\tau_k$, respectively. 
In this case, two PITE operators are summarized as
\begin{gather}
    \mathcal{U}_{\mathrm{PITE}}(\gamma_2,\Delta\tau_2) \mathcal{U}_{\mathrm{PITE}}(\gamma_1,\Delta\tau_1)
    \equiv
    \mathcal{U}_{\mathrm{two}}(\gamma_2, \gamma_1, \Delta\tau_2, \Delta\tau_1)
    \nonumber \\
    =
    \left(
        I_{2^{n}} \otimes W^{\dagger} \sigma_x R_{z}\left(-\frac{\pi}{2}\right)
    \right)
    \Big(
        e^{-i\kappa_2 \Theta_2} e^{i\kappa_1 \Theta_1}
        \otimes
        |0\rangle\langle 0| 
        + 
        e^{i\kappa_2 \Theta_2} e^{-i\kappa_1 \Theta_1}
        \otimes
        |1\rangle\langle 1| 
    \Big)
    \left(
        I_{2^{n}} \otimes WH
    \right) ,
\end{gather}
where the subscripts of $\Theta_k$ and $\kappa_k$ indicate that they are functions of $\gamma_k$ and $\Delta\tau_k$.
Furthermore, when the approximate circuits are adopted, the above equation can be rewritten as
\begin{gather}
    \mathcal{U}_{\mathrm{PITE}}^{(1)}(\theta_2,s_2\Delta\tau_2) \mathcal{U}_{\mathrm{PITE}}^{(1)}(\theta_1,s_1\Delta\tau_1)
    \equiv
    \mathcal{U}_{\mathrm{two}}^{(1)} (\theta_2, \theta_1, s_2\Delta\tau_2, s_1\Delta\tau_1)
    \nonumber \\
    =
    \left(
        I_{2^{n}} \otimes W^{\dagger}\sigma_x R_{z}\left(2(\theta_2-\theta_1)-\frac{\pi}{2}\right)
    \right)
    \nonumber \\
    \cdot
    \Big(
        U_{\mathrm{RTE}}^{\dagger}(s_2\Delta\tau_2)
        U_{\mathrm{RTE}}(s_1\Delta\tau_1)
        \otimes
        |0\rangle\langle 0| 
        + 
        U_{\mathrm{RTE}}(s_2\Delta\tau_2) 
        U_{\mathrm{RTE}}^{\dagger}(s_1\Delta\tau_1)
        \otimes
        |1\rangle\langle 1| 
    \Big)
    \left(
        I_{2^{n}} \otimes WH
    \right) .
\end{gather}
Here, the approximate PITE operator $\mathcal{U}_{\mathrm{PITE}}^{(1)}$ takes input parameters corresponding to $\gamma_k$ and $\Delta\tau_k$, 
but instead we adopt the equivalent but different input parameters $s_k\Delta\tau_k$ and $\theta_k$ for the approximate PITE as $\mathcal{U}_{\mathrm{PITE}}^{(1)}(s_k\Delta\tau_k, \theta_k)$. The reason of the transformation is to be easy to find the correspondence to the parameters appearing in the quantum circuit. 
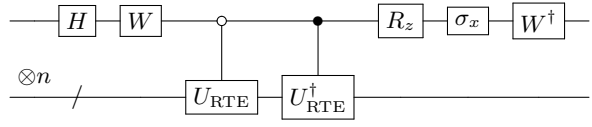
\begin{figure}
\centering
\mbox{ 
\Qcircuit @C=1em @R=1.5em{
\lstick{} & \qw                     & \gate{H}   & \gate{W} 
& \ctrlo{1}                                     & \ctrl{1}                                                & \gate{R_z} &\gate{\sigma_x} & \gate{W^{\dagger}} & \qw \\
\lstick{} & \ustick{\otimes n} \qw  & {/} \qw    & \qw      
& \gate{U_{\mathrm{RTE}}} & \gate{U_{\mathrm{RTE}}^{\dagger}} & \qw      & \qw          & \qw      & \qw \\
}
} 
\caption{
Quantum circuit for the two-step PITE operator represented as $\mathcal{U}_{\mathrm{two}}^{(1)} (\theta_2, \theta_1, s_2, s_1) \equiv \mathcal{U}_{\mathrm{PITE}}^{(1)}(\gamma_2,\Delta\tau_2) \mathcal{U}_{\mathrm{PITE}}^{(1)}(\gamma_1,\Delta\tau_1)$.
$U_{\mathrm{RTE}}\equiv U_{\mathrm{RTE}}(s_2 \Delta \tau_2 - s_1 \Delta \tau_1) $ and $R_z \equiv R_z(2(\theta_2-\theta_1)-\frac{\pi}{2})$ are used in this figure.
}
\label{circuit:imag_evol_as_part_of_real_evol:evolution_2nd_steps}
\end{figure}

As is the case with the pre-amplification operator, we can reduce the circuit depth by half with the use of relation, 
$
    U_{\mathrm{RTE}}(s_2\Delta\tau_2)
    U_{\mathrm{RTE}}^{\dagger}(s_1\Delta\tau_1) 
    = 
    U_{\mathrm{RTE}}(s_2\Delta\tau_2-s_1\Delta\tau_1)
$.
We show the quantum circuit for  $\mathcal{U}_{\mathrm{two}}^{(1)}$ in Fig. \ref{circuit:imag_evol_as_part_of_real_evol:evolution_2nd_steps}.
Especially, when $\gamma_{2} = \gamma_{1}$ and $\Delta\tau_2 = \Delta\tau_1$, $\mathcal{U}_{\mathrm{two}}^{(1)}$ becomes identity; $\mathcal{U}_{\mathrm{two}}^{(1)} (\theta_1, \theta_1, s_1\Delta\tau_1, s_1\Delta\tau_1)=I_{2^{n+1}}$.
However, we note that the success probability in this case becomes small in general because we cannot tune the parameter for PITE in the second step. 

Suppose the amplified state expressed in Eq. (\ref{eq:separable_state_by_pite}) is not a pure separable state but a somehow entangled state containing a small portion of a failure state. In that case, the failure state is contaminated with the success state in the second step. The failure state works as an imaginary-time evolution in a backward direction. We discuss the detail in Appendix \ref{estimation_of_error}.

\subsection{PITE combined with QAA in multiple steps}
\label{sec:circuit_PITEwQAAnstep}
In the previous subsection, we show that the separable state can be generated by PITE combined with QAA when we choose the optimal parameter $\gamma^{*}$ and repetition $m^{*}$.
After the PITE operator in the second step acts on the separable state, we can amplify the success probability of the state by using QAA in the same way. 
In other words, we use the unitary for producing the separable state as the reference circuit in the next step.
Thus, we can obtain the separable state again after the two steps of PITE operator.
Furthermore, by repeating this procedure, it is possible to obtain the state where the PITE operator acts multiple times as the separable state in principle.
In this subsection, we discuss the pre-amplification operator when we use PITE combined with QAA throughout all imaginary-time steps.

\subsubsection{Quantum circuit}
\begin{figure}
\centering
\mbox{ 
\Qcircuit @C=1em @R=1.5em{
\lstick{} & \qw                   & \qw     & \multigate{1}{\mathcal{U}_{\mathrm{PITE+ref}}^{[N_\tau]}} & \qw  &
\push{\rule{.5em}{.0em}\raisebox{-4.5em}{=}\rule{1.0em}{0em}} & 
\qw & \qw     & \multigate{1}{\widetilde{Q}_{N_\tau-1}^{m_{N_\tau-1}}} & 
\multigate{1}{\mathcal{U}_{\mathrm{PITE+ref}}^{[N_\tau-1]}} & \qw &
\multigate{1}{\mathcal{U}_{\mathrm{PITE}}}     & \qw 
\\
\lstick{} & \ustick{\otimes n} \qw & {/} \qw & \ghost{\mathcal{U}_{\mathrm{PITE+ref}}^{[N_\tau]}}        & \qw  &
\push{\rule{.5em}{.0em}\rule{1.0em}{0em}} & 
\ustick{\otimes n} \qw & {/}  \qw & \ghost{\widetilde{Q}_{N_\tau-1}^{m_{N_\tau-1}}}        & 
\ghost{\mathcal{U}_{\mathrm{PITE+ref}}^{[N_\tau-1]}}        & \qw &
\ghost{\mathcal{U}_{\mathrm{PITE}}}           & \qw   
\gategroup{1}{9}{2}{10}{.9em}{--}  \\
\lstick{} & & & & & & & & &\push{\raisebox{.5em}{=$U_{\mathrm{ref}}^{[N_\tau]}$}}  & & &
}
} 
\caption{
Schematic of quantum circuit generating $N_{\tau}$th step of PITE in separable form with QAA by using the separable state after the $(N_{\tau}-1)$ steps.
}
\label{circuit:u_pite_in_ntau_step}
\end{figure}
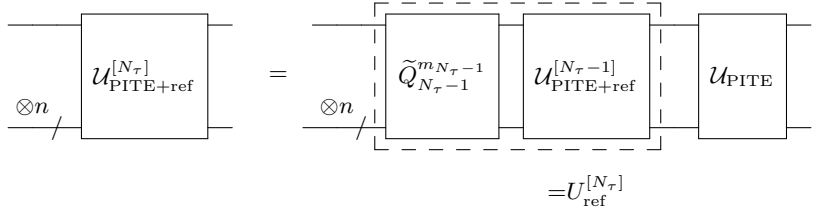

Let us consider a reference circuit $U_{\mathrm{ref}}^{[N_{\tau}]}$ in the $N_{\tau}$th step, which generates a separable state as 
$
    U_{\mathrm{ref}}^{[N_{\tau}]} |0\rangle_{n} \otimes |0\rangle
    =
    e^{-(N_{\tau}-1) \Delta\tau \mathcal{H}} |\psi\rangle \otimes |0\rangle
$.
The reference circuit in the first step is set to $U_{\mathrm{ref}}^{[1]} = U_{\mathrm{ref}} \otimes I_{2}$.
The way to construct the reference circuit $U_{\mathrm{ref}}^{[N_{\tau}]}$ can be derived inductively.
We consider the  PITE operator on the state generated by the reference circuit $U_{\mathrm{ref}}^{[N_{\tau}]}$ in the $N_{\tau}$th step. 
This process is expressed as
\begin{gather}
    \mathcal{U}_{\mathrm{PITE+ref}}^{[N_{\tau}]}
    =
    \mathcal{U}_{\mathrm{PITE}}
    U_{\mathrm{ref}}^{[N_{\tau}]} .
\end{gather}
The operation of the unitary $\mathcal{U}_{\mathrm{PITE+ref}}^{[N_{\tau}]}$ leads to entanglement of the desired state 
$
    e^{-N_{\tau} \Delta\tau \mathcal{H}} |\psi\rangle \otimes |0\rangle
$
and the orthogonal state. 
By using the QAA framework as desicribed in Sec.\ref{sec:circuit_dite}, we can get the desired state 
$
    e^{-N_{\tau} \Delta\tau \mathcal{H}} |\psi\rangle \otimes |0\rangle
$
as a separable state. 
Let $m_k$ be a number of repetitions of the pre-amplification operator $\widetilde{Q}_{k}$ in the $k$th step. 
In the same way as the first step, the separable state is obtained by following the unitary, which is equal to the reference circuit of the $(N_{\tau}+1)$th step,
\begin{gather}
    U_{\mathrm{ref}}^{[N_\tau+1]}
    = 
    \mathcal{U}_{\mathrm{PITE+ref}}^{[N_{\tau}]}
    \widetilde{Q}_{N_{\tau}}^{m_{N_{\tau}}^{*}} .
\end{gather}
In this case, the reference unitary $U_{\mathrm{ref}}^{[N_\tau+1]}$ acts on the ($n+1$)-qubit system including the ancillary qubit.
The reference circuit in any step can be obtained by recursively applying this procedure.
The quantum circuit for the unitary $\mathcal{U}_{\mathrm{PITE+ref}}^{[N_{\tau}]}$
 is shown in Fig. \ref{circuit:u_pite_in_ntau_step}.

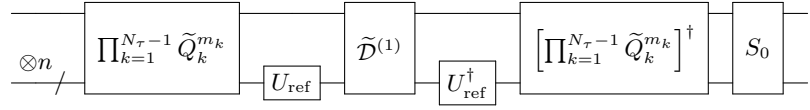
\begin{figure}
\centering
\mbox{ 
\Qcircuit @C=1em @R=1.5em{
\lstick{} & \qw & \qw     
& \multigate{1}{\prod_{k=1}^{N_\tau-1} \widetilde{Q}_{k}^{m_k}}
& \qw & \multigate{1}{\widetilde{\mathcal{D}}^{(1)}} & \qw 
& \multigate{1}{\left[\prod_{k=1}^{N_\tau-1} \widetilde{Q}_{k}^{m_k}\right]^{\dagger}}
& \multigate{1}{S_0} & \qw 
\\
\lstick{} & \ustick{\otimes n} \qw & {/} \qw 
& \ghost{\prod_{k=1}^{N_\tau-1} \widetilde{Q}_i^{m_k}}
& \gate{U_{\mathrm{ref}}} & \ghost{\widetilde{\mathcal{D}}^{(1)}}   & \gate{U_{\mathrm{ref}}^{\dagger}} 
& \ghost{\left[\prod_{k=1}^{N_\tau-1} \widetilde{Q}_k^{m_k}\right]^{\dagger}}
& \ghost{S_0}        & \qw \\
}
} 
\caption{
Quantum circuit for the pre-amplification operator $\widetilde{Q}_{N_{\tau}}$ in $N_\tau$th step.
}
\label{circuit:imag_evol_as_part_of_real_evol:circuit_e}
\end{figure}

\begin{figure}
\centering
\mbox{ 
\Qcircuit @C=1em @R=1.5em{
\lstick{} & \qw & \gate{H}                & \gate{W} & \ctrlo{1}                                     & \ctrl{1}                                                & \gate{R_z} &\gate{\sigma_y} & \gate{W^{\dagger}} & \gate{H} &\qw \\
\lstick{} & \ustick{\otimes n} \qw & {/} \qw & \qw      & \gate{U^2_{\mathrm{RTE}}} & \gate{U_{\mathrm{RTE}}^{\dagger 2}} & \qw      & \qw          & \qw      & \qw & \qw \\
}
} 
\caption{
Quantum circuit for $\mathcal{D}_{\mathrm{two}}^{(1)}(\gamma_2, \gamma_1, \Delta\tau_2, \Delta\tau_1)$. 
$U^2_{\mathrm{RTE}} = U^2_{\mathrm{RTE}}(s_2 \Delta \tau_2 - s_1 \Delta \tau_1)$ and $R_z = R_z(-4(\theta_2-\theta_1))$ are used in this figure.
}
\label{circuit:imag_evol_as_part_of_real_evol:circuit_D_two}
\end{figure}
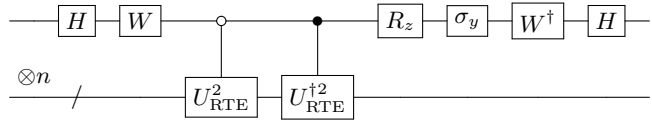

Below we describe the concrete expressions for the unitary $\mathcal{U}_{\mathrm{PITE+ref}}^{[N_{\tau}]}$ and the pre-amplification operator $\widetilde{Q}_{N_\tau}$ in the $N_{\tau}$th step.
The unitary $\mathcal{U}_{\mathrm{PITE+ref}}^{[N_{\tau}]}$ representing PITE combined with QAA in the $N_{\tau}$th step is expanded in a chain as 
\begin{gather}
    \mathcal{U}_{\mathrm{PITE+ref}}^{[N_{\tau}]}
    =
    \mathcal{U}_{\mathrm{PITE}}(\gamma_{N_{\tau}}^{*}, \Delta\tau_{N_{\tau}} )
    U_{\mathrm{ref}}^{[N_\tau]}
    \nonumber \\
    = 
    \left[
        \prod_{k=N_{\tau}}^{1}
        \mathcal{U}_{\mathrm{PITE}}(\gamma_{k}^{*}, \Delta\tau_k)
    \right]
    \Big(U_{\mathrm{ref}} \otimes I_2 \Big)
    \left[
        \prod_{k=1}^{N_{\tau}-1}
        \widetilde{Q}_{k}^{m_{k}^{*}} 
    \right] .
\end{gather}
For practical purposes, we adopt the approximate circuit $\mathcal{C}^{(1)}_{\mathrm{PITE}}$ below, but its extension to the general case is straightforward.
The product of the series of the PITE operators
$
    \prod_{k=N_{\tau}}^{1} \mathcal{U}_{\mathrm{PITE}}(\gamma_{k}, \Delta\tau_k) 
    = 
    \mathcal{U}_{\mathrm{PITE}}(\gamma_{N_{\tau}}, \Delta\tau_{n_\tau1})
    \cdots 
    \mathcal{U}_{\mathrm{PITE}}(\gamma_{1}, \Delta\tau_{1})
$ 
is rewritten with the approximate circuits as
\begin{gather}
\begin{dcases}
    \mathcal{U}_{\mathrm{PITE}}^{(1)}
    \left(
        \sum_{i=1}^{N_\tau} (-1)^{k+1} \theta_k, \sum_{k=1}^{N_\tau} (-1)^{k+1} s_k\Delta\tau_k
    \right)
    &
    \text{for $N_{\tau}$ odd}
    \\
    \mathcal{U}_{\mathrm{two}}^{(1)}
    \left(
        \theta_{N_\tau},
        \sum_{k=1}^{N_{\tau}-1} (-1)^{k+1} \theta_k, 
        s_{N_\tau}\Delta\tau_{N_{\tau}} ,
        \sum_{k=1}^{N_{\tau}-1} (-1)^{k+1} s_k \Delta\tau_k
    \right)
    &
    \text{for $N_{\tau}$ even} ,
\end{dcases}
\end{gather}
where the approximate unitary $\prod_{k=N_\tau}^{1} \mathcal{U}_{\mathrm{PITE}}^{(1)}(\theta_k, s_k\Delta\tau_k)$ is implemented by $\mathcal{U}_{\mathrm{PITE}}^{(1)}$ or $\mathcal{U}_{\mathrm{two}}^{(1)}$ depending on odd or even numbers of $N_\tau$, respectively. In the above equation, we replace the input parameters from $(\gamma_{k}, \Delta\tau_k)$ with $(\theta_k, s_k\Delta\tau_k)$, because of the same reasons in subsection~{\ref{sec:circuit_dite}}.
Using the above relations, the pre-amplification operator 
$
    \widetilde{Q}_{N_\tau}
    \equiv
    -S_{0} 
    \mathcal{U}_{\mathrm{PITE+ref}}^{[N_\tau] \dagger}
    S_{\chi}
    \mathcal{U}_{\mathrm{PITE+ref}}^{[N_\tau]}
$
in the $N_\tau$th step is expressed as 
\begin{gather}
    \widetilde{Q}_{N_{\tau}}
    =
    S_{0}
    \left[
        \prod_{k=1}^{N_{\tau}-1}
        \widetilde{Q}_{k}^{m_{k}^{*}}
    \right]^{\dagger}
    \Big( 
        U_{\mathrm{ref}}^{\dagger} \otimes I_2 
    \Big)
    \widetilde{\mathcal{D}}^{(1)}
    \left(
        \boldsymbol{\theta},
        \boldsymbol{s},
        \boldsymbol{\Delta \tau}
    \right)
    \Big( 
        U_{\mathrm{ref}} \otimes I_2
    \Big)
    \left[
        \prod_{k=1}^{N_{\tau}-1}
        \widetilde{Q}_{k}^{m_{k}^{*}}
    \right],
\end{gather}
where the quantum circuit for $\widetilde{Q}_{N_{\tau}}$ is presented in Fig. \ref{circuit:imag_evol_as_part_of_real_evol:circuit_e} and the unitary
$
    \widetilde{\mathcal{D}}^{(1)}
    \left(
        \boldsymbol{\theta},
        \boldsymbol{s},
        \boldsymbol{\Delta \tau}
    \right)
$ 
is defined as
\begin{gather}
    \begin{dcases}
    \mathcal{D}^{(1)}
    \left(
        \sum_{k=1}^{N_{\tau}} (-1)^{k+1} \theta_k, \sum_{k=1}^{N_{\tau}} (-1)^{k+1} s_k \Delta\tau_k
    \right)
    &
    \text{for $N_{\tau}$ odd}
    \\
    \mathcal{D}_{\mathrm{two}}^{(1)}
    \left(
        \theta_{N_\tau},
        \sum_{k=1}^{N_\tau-1} (-1)^{k+1} \theta_k, 
        s_{N_\tau} \Delta \tau_{N_\tau},
        \sum_{k=1}^{N_{\tau}-1} (-1)^{k+1} s_k \Delta\tau_k
    \right)
    &
    \text{for $N_{\tau}$ even} ,
    \end{dcases} 
\end{gather}
where $\boldsymbol{\theta} = (\theta_1,\ldots, \theta_{N_{\tau}})$, $\boldsymbol{s} = (s_1,\ldots, s_{N_{\tau}})$, and $\boldsymbol{\Delta \tau} = (\Delta \tau_1,\ldots, \Delta \tau_{N_{\tau}})$.
Here, the unitary $\mathcal{D}_{\mathrm{two}}$ is defined in the same way as $\mathcal{D}$, 
\begin{gather}
    \mathcal{D}_{\mathrm{two}}(\gamma_2, \gamma_1, \Delta\tau_2, \Delta\tau_1)
    \equiv
    -\mathcal{U}_{\mathrm{two}}^{\dagger}(\gamma_2, \gamma_1, \Delta\tau_2, \Delta\tau_1)
    S_{\chi}
    \mathcal{U}_{\mathrm{two}}(\gamma_2, \gamma_1, \Delta\tau_2, \Delta\tau_1)
    .
\end{gather}
The concrete expression of $\mathcal{D}_{\mathrm{two}}$ with the use of the approximate PITE is derived as 
\begin{gather}
	\mathcal{D}_{\mathrm{two}}^{(1)}(\gamma_2, \gamma_1, s_2\Delta\tau_2, s_1\Delta\tau_1)
    \equiv
    -\mathcal{U}_{\mathrm{two}}^{(1) \dagger}(\theta_2,\theta_1, s_2\Delta\tau_2, s_1\Delta\tau_1)
    S_{\chi}
    \mathcal{U}_{\mathrm{two}}^{(1)}(\theta_2,\theta_1, s_2\Delta\tau_2, s_1\Delta\tau_1)
    \nonumber \\
    =
    \Big(
        I_{2^{n}} \otimes H W^{\dagger} \sigma_y R_z\big(4(\theta_2-\theta_1)\big)
    \Big)
    \nonumber \\
    \cdot
    \Big(
        U_{\mathrm{RTE}}^2(s_2\Delta\tau_2 - s_1\Delta\tau_1) \otimes |0\rangle \langle 0|
        +
        U_{\mathrm{RTE}}^{\dagger 2}(s_2\Delta\tau_2 - s_1\Delta\tau_1) \otimes |1\rangle \langle 1|
    \Big)
    \left(
        I_{2^{n}} \otimes WH
    \right) ,
\end{gather}
where the corresponding circuit is depicted in Fig \ref{circuit:imag_evol_as_part_of_real_evol:circuit_D_two}. 

\subsubsection{Computational cost}
Let us discuss the computational cost for the pre-amplification operator $\widetilde{Q}_{N_\tau}$ in the $N_\tau$th step.
Because the computational cost for $\mathcal{D}^{(1)}$ and $\mathcal{D}_{\mathrm{two}}^{(1)}$ is almost identical to the one for $\mathcal{U}_{\mathrm{PITE}}^{(1)}$,
the computational cost is estimated as
\begin{gather}
\begin{aligned}
    c_{\widetilde{Q}_{N_\tau}}
    &\approx
    c_{S_0}
    +
    c_{\mathcal{U}_{\mathrm{PITE}}}
    +
    2 c_{U_{\mathrm{ref}}}
    +
    2\sum_{k=1}^{N_\tau-1}
    m_k^{*} c_{\widetilde{Q}_{k}}
    \nonumber \\
    &=
    (1 + 2m_{N_\tau-1}^{*})
    \left(
        c_{S_0}
        +
        c_{\mathcal{U}_{\mathrm{PITE}}}
        +
        2 c_{U_{\mathrm{ref}}}
        +
         2\sum_{k=1}^{N_\tau-2}
        m_k^{*} c_{\widetilde{Q}_{k}}
    \right)
    \nonumber \\
    &=
    \left(
        c_{S_0}
        +
        c_{\mathcal{U}_{\mathrm{PITE}}}
        +
        2 c_{U_{\mathrm{ref}}}
    \right)
    \prod_{k=N_\tau-1}^{1}
    (1 + 2m_{k}^{*})
    ,
\end{aligned}
\end{gather}
where we use $c_{\widetilde{Q}_{1}} = c_{\widetilde{Q}} = c_{S_0} + c_{\mathcal{U}_{\mathrm{PITE}}} + 2 c_{U_{\mathrm{ref}}}$.
This means that even if $m_{k}^{*}=1$, PITE with QAA for $N_\tau$ steps requires $3^{N_\tau-1}(c_{S_0} + c_{\mathcal{U}_{\mathrm{PITE}}} + 2 c_{U_{\mathrm{ref}}})$ computational cost.
In other words, the computational cost for the pre-amplification operator increases power with respect to the number of PITE steps $N_{\tau}$.
In addition, the computational cost for $\mathcal{U}_{\mathrm{PITE}}^{[N_{\tau}]}$ is calculated as
\begin{gather}
\begin{aligned}
    c_{\mathcal{U}_{\mathrm{PITE}}^{[N_{\tau}]}}
    &\approx
    c_{\mathcal{U}_{\mathrm{PITE}}}
    +
    c_{U_{\mathrm{ref}}}
    +
    \sum_{k=1}^{N_{\tau}-1}
    m_k^{*} c_{\widetilde{Q}_{k}}
    \nonumber \\
    &=
    c_{\mathcal{U}_{\mathrm{PITE}}}
    +
    c_{U_{\mathrm{ref}}}
    +
    \left(
        c_{S_0}
        +
        c_{\mathcal{U}_{\mathrm{PITE}}}
        +
        2 c_{U_{\mathrm{ref}}}
    \right)
    \sum_{k=1}^{N_\tau-1}
    m_k^{*}
    \left[
        \prod_{j=1}^{j=k-1}
        (1 + 2m_{j}^{*})
    \right] .
\end{aligned}
\end{gather}
Similarly, even when $m_{k}^{*}=1$, the computational cost increases the power;
\begin{gather}
    c_{\mathcal{U}_{\mathrm{PITE}}^{[N_{\tau}]}}
    \approx c_{\mathrm{ITE}}
    +
    c_{U_{\mathrm{ref}}}
    +
    \frac{1}{2}
    \left(
        c_{S_0}
        +
        c_{\mathcal{U}_{\mathrm{PITE}}}
        +
        2 c_{U_{\mathrm{ref}}}
    \right)(3^{N_\tau-1}-1) .
\end{gather}
This increase in power is not preferable. 
However, the success probability increases monotonically as the number of steps increases.
This behavior is proven in the case of a two-level system \cite{PITE}, but this proof might be extended to general cases.
It is because the high energy state decreases as the imaginary-time step evolves, and the smaller decaying the high energy state makes the normalization constant $\alpha^2 = \langle\psi |e^{-2\Delta\tau\mathcal{H}}| \psi \rangle$ smaller.   
Therefore, adopting the method as introduced in this subsection is desirable for a few ITE steps where the success probability is small.  
In another way, let us consider that only one imaginary-time step uses the combination of PITE and QAA.
In this way, first, we estimate the output state of PITE by using quantum state tomography \cite{Nielsen2000Book}, from which we reconstruct the reference circuit, $U_{\mathrm{ref}}$ \cite{Shende2005ACM}, in the next imaginary-time step. 
By repeatedly proceeding with these two techniques, we achieve the ground state.

\section{Results}
\label{sec:results}
In this study, we use the first-order approximate circuit $\mathcal{C}_{\mathrm{PITE}}^{(1)}$ for numerical simulation in Fig. \ref{circuit:imag_evol_as_part_of_real_evol:evolution_1st_order}. 
Also, we use Qiskit \cite{Qiskit} for quantum circuit simulation.

\subsection{Second-quantized Hamiltonian}
\label{sec:results_2ndHamil}

In this section, we consider the max-cut problem as an example.
The max-cut problem is one of the most famous combinatorial optimization problems.
The Hamiltonian for the max-cut problem in the qubit representation, given as an Ising Hamiltonian, is expressed as \cite{kirkpatrick1983optimization,Farhi2014arxiv}
\begin{gather}
    \mathcal{H} 
    = 
    - \sum_{(i,j)\in E} d_{i,j} \frac{1-\sigma_{z,i}\otimes \sigma_{z,j}}{2} ,
\end{gather}
where $E$ is the set of edges in a graph and $d_{i,j}$ is the weight of edge $(i,j)$. 
The problem with seeking the ground state of Ising spin glass is known as NP-hard \cite{garey1974some,Barahona_1982}.
The expressions of other representative combinatorial optimization problems in the qubit representation are presented in reference \cite{lucas2014ising}.

\subsubsection{Computational cost}
\label{sec:results_2ndHamil_cost}
\begin{figure}[ht]
    \centering
    \includegraphics[width=1.0 \textwidth]{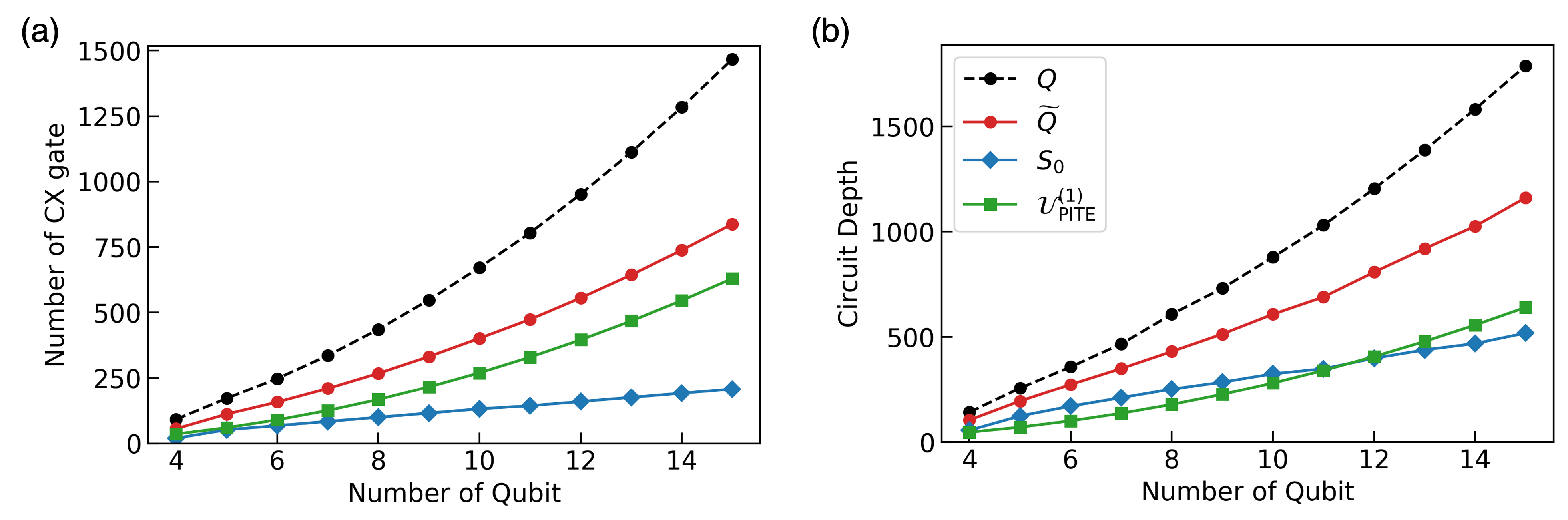}
 \caption{
The dependence of the number of CNOT gates, or equivalently CX gates, (a) and the circuit depth (b) depending on the number of qubits for a step of the PITE operation applied to the fully-connected Ising Hamiltonian. The computational costs for the amplification operator $Q$ (black), the pre-amplification operator $\widetilde{Q}$ (red), the zero reflection $S_0$ (blue), and the approximate PITE circuit $\mathcal{U}_{\mathrm{PITE}}^{(1)}$ (green) are presented. 
 }
\label{fig:pite_qaa_cost}
\end{figure}

The circuit depth and the number of CNOT gates for the pre-amplification operator $\widetilde{Q}$ are shown in figures \ref{fig:pite_qaa_cost}(a) and (b), respectively.
The pre-amplification operator includes the controlled-RTE operators of a given Hamiltonian.
This numerical simulation uses the Hamiltonian of fully-connected graphs.
Here, we adopt a superposition of the computational bases with equal weights as an initial state. Thus, the reference circuit for producing the initial state is $U_{\text{ref}}=H^{\otimes n}$ with $H$ being the Hadamard gate.
The zero reflection for $n$-qubit system can be implemented from a multiply controlled Toffoli gate $C^{n-1}(\sigma_x)$
and the Pauli-$X$ gates as $S_0=- \sigma_x^{\otimes n} (I_{2^{n-1}}\otimes H) C^{n-1}(\sigma_x) (I_{2^{n-1}}\otimes H) \sigma_x^{\otimes n}$. 
The multiply controlled Toffoli gate $C^{n}(\sigma_x)$ requires $\mathcal{O}(n^2)$ CNOT gates for the circuit implementation. 
It is allowed to implement the multiply controlled Toffoli gate $C^{n}(\sigma_x)$ with $\mathcal{O}(n)$ CNOT gates through the use of an ancillary qubit \cite{AdrianoPRA1995}. 
In this paper, we realize the multiply controlled Toffoli gate by combining the implementation method utilized with one ancillary qubit (\cite{AdrianoPRA1995} Corollary 7.4) and a relative-phase Toffoli gate presented in \cite{MaslovPRA2016}.

First, we discuss the dependence of the number of CNOT gates on the number of qubits in Fig. \ref{fig:pite_qaa_cost}(a).
As mentioned above, we recognize the linearity with respect to the number of CNOT gates for the zero reflection.
Specifically, the number of CNOT gates for the zero reflection increases as relation, $n_{\text{cnot}}= 16(n-3)$ in the range $n \geq 11$ (see Ref. \cite{AdrianoPRA1995} and \cite{MaslovPRA2016}). 
The quadratic increase in the number of terms for the fully-connected Ising Hamiltonian leads to a parabolic shape in PITE, as depicted with the green line. 
By translating from the amplification operator $Q$ to the pre-amplification operator $\widetilde{Q}$, we see the reduction of CNOT gates by about the quantum circuit for $\mathcal{U}_{\mathrm{PITE}}^{(1)}$.

Next, we mention the dependence of the circuit depth on the number of qubits in Fig. \ref{fig:pite_qaa_cost} (b).
The circuit depth for the zero reflection is also linear with the number of qubits \cite{MaslovPRA2016,he2017decompositions}. 
As a more sophisticated method, reference \cite{he2017decompositions} proposed $\mathcal{O}(\log n)$ circuit depth implementation method, which is realized by utilizing $n$ ancillary qubits and disentanglement operation (measurement and corresponding recovery operation). 
The circuit depth for the approximate PITE circuit $\mathcal{U}_{\mathrm{PITE}}^{(1)}$ also increases quadratically, as in the case of the CNOT gates. 
Furthermore, we see the reduction of the circuit depth by translating from the amplification operator $Q$ to the pre-amplification operator $\widetilde{Q}$.
These results are consistent with the statement in the section \ref{section:QITE_with_QAA_operation_cost} and convincing the effectiveness of our technique.

\subsubsection{Success probability}
\label{sec:results_2ndHamil_prob}
\begin{figure}[ht]
        \centering
        \includegraphics[width=1.0 \textwidth]{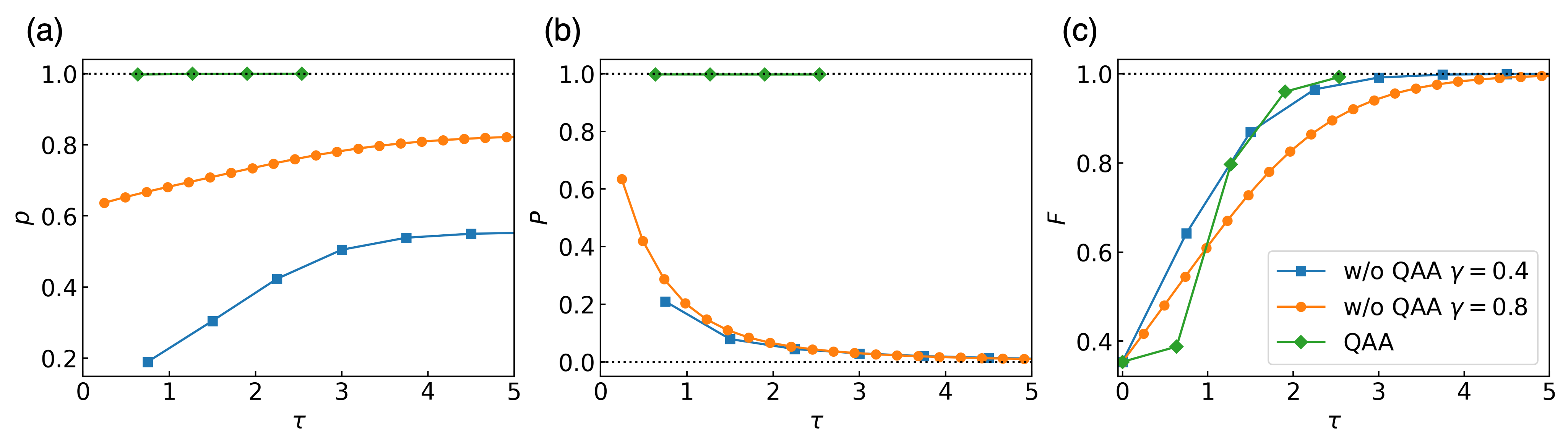}
        \label{fig:pite_nstep}
 \caption{
    The simulation result of the max-cut problem with four qubits for probability $p_k$ (a), total probability $P_k=\prod_{\ell=1}^{k} p_{\ell} $ (b), and fidelity $F$ (c) depending on imaginary-time $\tau$. Blue and orange lines show the results solely performed by PITE with $\gamma=0.4, 0.8$, respectively. Green lines represent the results by PITE combined with QAA. Each imaginary-time step is drawn with a point.
 }
 \label{fig:pite_result}
\end{figure}

Here, as a computational target, we consider the max-cut problem comprised of four qubits, which has edges $E=\{(0,1),(1,2),(2,3),(3,0),(0,2)\}$ with  weight $d_{i,j}=1$.
The solution in this graph is $| \phi_{\mathrm{gs}} \rangle =  (|0101\rangle+|1010\rangle)/\sqrt{2}$
for the eigenstate and $\lambda_{\min}=-4$ for eigenvalues.
We plot the success probabilities, total success probabilities (success probability throughout all imaginary time steps expressed as $P_k=\prod_{\ell=1}^{k} p_{\ell} $, where the $p_{\ell}$ represents the probability of obtaining the success state at $l$-th step), and fidelities,
$
    F = |\langle \psi(\tau_k)|\phi_{\mathrm{gs}}\rangle|^{2}
$
, until reaching the ground state in Figure \ref{fig:pite_result} (a), (b), and (c), respectively.
As the time step size is set to a constant $\Delta \tau$, the imaginary time at the $k$th step is expressed as $\tau_k = k\Delta\tau$.

In the numerical simulation, we use different $\Delta\tau$ depending on the PITE parameter $\gamma$ because the rotation angle of the RTE operator is proportional to the product $s\Delta\tau$. 
The determination method for $\Delta\tau$ is mentioned in the appendix \ref{appendix:other_parameter_determination}.
We use $\Delta\tau = 0.75$ and $0.25$ for PITE with the parameters $\gamma = 0.4$ and $0.8$, respectively.  
With the repetitions $m^{*}=1$ for QAA, the optimized parameter and the step size are calculated as $\gamma^{*}=0.46$ and $\Delta\tau = 0.63$, respectively.
First, let us look at the probabilities shown in Fig. \ref{fig:pite_result} (a) and (b).
When QAA is not used, the success probabilities increase monotonically and finally converge to a certain constant value.
We also confirm that the larger the parameter $\gamma$ is, the greater the success probability is, as shown in $\mathbb{P}_{0} = \gamma^2 \alpha^2$.
This is a reasonable fact already stated in the previous studies \cite{PITE} for a mathematical discussion.
In addition, the total success probabilities in Fig. \ref{fig:pite_result} (b) are monotonically decreasing. Then the total success probability is almost zero when we get the ground state.
On the other hand, when we use PITE combined with QAA, both the success probability and the total success probability are one within the numerical error, from which we can recognize that QAA dramatically improves the success probability.
The convergence to the ground state is confirmed with all conditions in Fig. \ref{fig:pite_result} (c).

Although the computational cost would increase by power of the imaginary time steps necessary when PITE and QAA were used together, we have found that convergence to the exact solution was achieved with a small number of imaginary time steps because the larger $\Delta\tau$ is permitted due to small $\gamma$ as a secondary benefit of the QAA. 
The circuit depth and the number of CNOT gates in this simulation were (68,  200,  596, 1784.) and (108, 346, 1052, 3178) according to PITE steps, respectively.
In contrast, the circuit depth and the number of CNOT gates of the approximate PITE circuit without QAA are 24 and 26, respectively.

We used the Hadamard gates in a reference circuit to prepare the superposition with an equal probability for all the computational bases. However, as discussed in Sec. \ref{sec:circuit_dite}, 
the superposition with an equal probability may occur in the exponential decay of the normalization constant $\alpha = \sqrt{\langle\psi | e^{-2\Delta\tau \mathcal{H}} |\psi\rangle}$ as the number of qubits increases. 
One possibility is to start a better initial state to overcome the shortcoming, e.g.,  using an approximate solution yielded by another algorithm such as a warm-start strategy actively researched in the field of quantum approximate optimization algorithm (QAOA) \cite{Farhi2000arXiv}, which improves the performance by starting from the approximately computed initial state \cite{Egger2021Quantum,Tate2020arXiv,Okada2022arXiv}.

\subsection{First-quantized Hamiltonian}
\label{sec:results_1stHamil}
An approach to exploit the threatening potential of a quantum computer by employing an encoding of spatial grids was first proposed by Zalka and Wiesner \cite{Zalka1998, Wiesner1996arXiv}.
Kassal et al. realized quantum circuits for real-time evolution based on the first-quantized Hamiltonian by using a phase-kickback technique to implement potential terms \cite{Kassal2008PNAS}. 
Later, many developments are presented for efficient constructing circuits, including truncated Taylor series algorithm \cite{Kivlichan2017JPA} dual representation \cite{Babbush2018PRX}, interaction-picture \cite{Babbush2019npjQI, Su2021PRXQuantum}, and qubitization \cite{Su2021PRXQuantum}.
Besides, the first-quantized-based (grid-based) approach also shows several advantages over the second-quantized Hamiltonian in the terms of the variational quantum algorithm, which requires measuring only momentum and positional space \cite{Ollitrault2022arXiv}. 

Let us consider a one-electron confined in a one-dimensional harmonic potential for demonstration of our method for electronic structure calculations. The first-quantized Hamiltonian in this situation is expressed as, 
\begin{gather}
    \mathcal{H} = 
    \frac{p^2}{2m} 
    + 
    \frac{m\omega^2}{2}\left(x-\frac{L}{2}\right)^2
\end{gather}
where $m=1$ is the electron mass and $\omega=1$ is the frequency of the confining potential. We adopt $L=14$ in this simulation. We discretize the space into $N=2^n$ with an equal mesh, and the wave function of the electron is encoded on the space meshes.

\subsubsection{Computational cost}
\label{sec:results_1stHamil_cost}
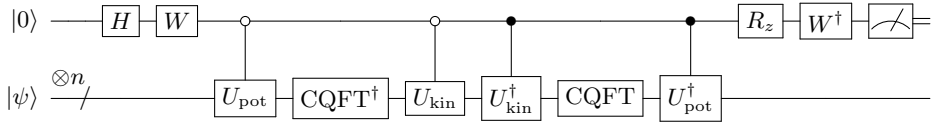
\begin{figure}[ht]
\centering
\mbox{
\Qcircuit @C=.7em @R=1.5em { 
\lstick{\ | 0 \rangle}  & \qw                    & \qw & \gate{H} & \gate{W} & \ctrlo{1} & \qw & \ctrlo{1} & \ctrl{1} & \qw     & \ctrl{1} & \gate{R_z} & \gate{W^\dagger} & \meter & \cw \\
\lstick{\ | \psi \rangle} & \ustick{\otimes n} \qw & {/} \qw  & \qw & \qw      & \gate{U_{\mathrm{pot}}} & \gate{\mathrm{CQFT}^{\dagger}} & \gate{U_{\mathrm{kin}}} & \gate{U_{\mathrm{kin}}^{\dagger}} & \gate{\mathrm{CQFT}} & \gate{U_{\mathrm{pot}}^\dagger} & \qw & \qw & \qw & \qw \\
} 
}
\caption{
Quantum circuit $\mathcal{C}_{\mathrm{PITE}}^{(1)}$ for approximate PITE operator 
for the first-quantized Hamiltonian. The potential-phase gate $U_{\mathrm{pot}}$ for the potential-evolution operator $e^{-i\hat{V}s\Delta\tau}$ and the kinetic-phase gate $U_{\mathrm{kine}}$ for the kinetic-evolution operator $e^{-i\hat{T}s\Delta\tau}$ are involved.
}
\label{circuit:imag_evol_as_part_of_real_evol:first_quantization}
\end{figure}
We explain a quantum circuit for simulating the first-quantized Hamiltonian first.
To implement the real-time evolution in the approximate imaginary time evolution method based on the first-quantized Hamiltonian, we divide the Hamiltonian into the kinetic part and potential part as $\mathcal{H} = \hat{T} + \hat{V}$, and the first order Suzuki-Trotter decomposition is performed as
$
    e^{i s\Delta \tau(\hat{T}+\hat{V})} 
    = 
    e^{i s\Delta \tau \hat{T}} 
    e^{i s\Delta \tau \hat{V}}
    + 
    \mathcal{O}({\Delta \tau}^2)
$. 
A potential-phase gate $U_{\mathrm{pot}}$ for the potential-evolution operator $e^{-i\hat{V}(x)t}$ acts on the position eigenstates diagonally as $U_{\mathrm{pot}}(t)|j\rangle_{n} \equiv \exp[-i\hat{V}(x^{(j)})t]|j\rangle_n$.
We use the technique \cite{Ollitrault2020PRL,Benenti2008AJP} for implementing a diagonal matrix given as $p$th order polynomial of position using several controlled-phase gates with $\mathcal{O}(n^{p})$.
To implement the kinetic part, we perform centered quantum Fourier transformation (CQFT), \cite{Somma2015arXiv, Ollitrault2020PRL} which shifts the center of momentum space to zero. 
Thus, the kinetic operator becomes diagonal unitary for the computational basis and can be implemented in the same way as the potential part.
The initial state is prepared as the superposition of the ground-state and three lowest excited states in this simulation, $|\psi\rangle = (|\phi_{\mathrm{gs}}\rangle + |\phi_{\mathrm{es1}}\rangle + |\phi_{\mathrm{es2}}\rangle + |\phi_{\mathrm{es3}}\rangle)/2$ \cite{PITE}. 
For preparing such initial state, we use the recursive construction method proposed in \cite{Shende2005ACM}. 
This synthesizing technique requires $2^n-2$ CNOT gates for realizing the state preparation.  
The implementation of the zero reflection is the same as described in the previous section \ref{sec:results_1stHamil_cost}. 

\begin{figure}[ht]
    \centering
    \includegraphics[width=1.0 \textwidth]{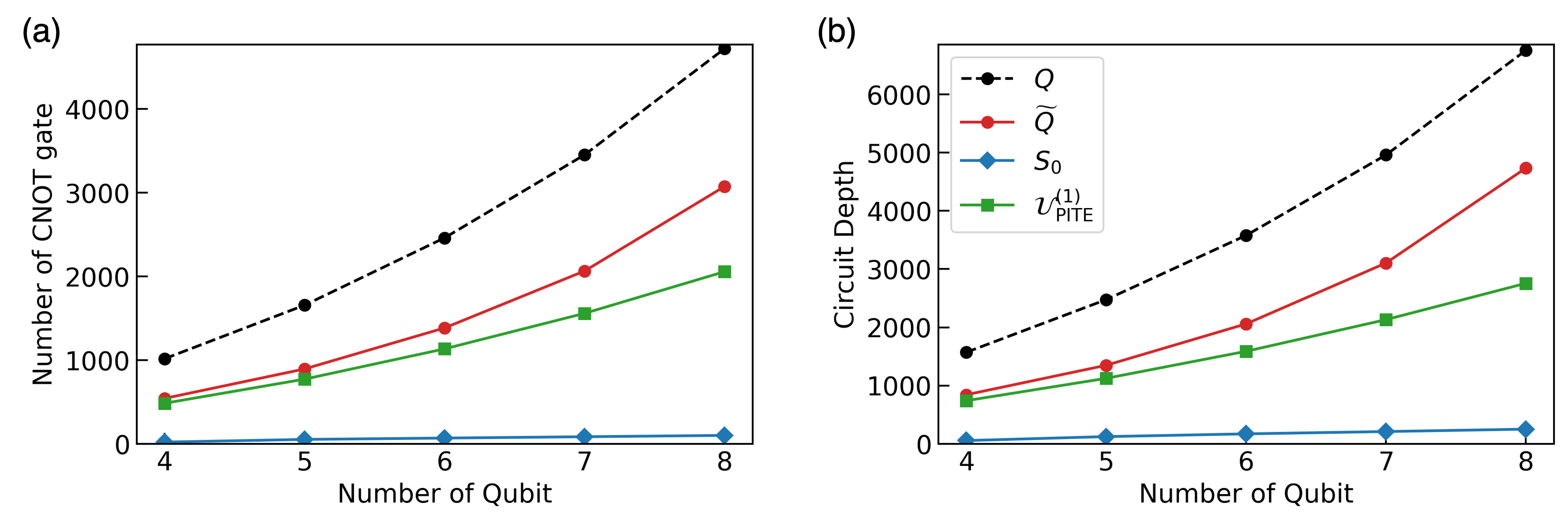}
 \caption{
 The dependence of the number of CNOT gates (a) and the circuit depth (b) depending on the number of qubits for  a step of the PITE operation applied to one electron confined in a one-dimensional harmonic potential. The computational costs for the amplification operator $Q$ (black), the pre-amplification operator $\widetilde{Q}$ (red), the zero reflection $S_0$ (blue), and the approximate PITE circuit $\mathcal{U}_{\mathrm{PITE}}^{(1)}$ (green) are presented. 
 }
 \label{fig:pite_qaa_cost_fqh}
\end{figure}

We depict the circuit depth and the number of CNOT gates for pre-amplification operator $\widetilde{Q}$ in figures \ref{fig:pite_qaa_cost_fqh} (a) and (b), respectively.
First, we confirmed the reduction by replacing the amplification operator with the pre-amplification operator, which is the benefit of this work. 
Linear-scaling of increase of the zero reflection with the number of qubits is the same as stated  in figure~\ref{fig:pite_qaa_cost}. 
The approximate PITE circuit is mainly composed of two CQFT and four phase gates. This computational cost is analytically estimated as $O(n^2)$. 
The PITE with QAA requires more CNOT gates and deeper circuit depth even for the small qubits as shown in Fig. \ref{fig:pite_qaa_cost_fqh} compared with VQE methods, but yields benefits in the rage of larger number of qubits. 
The CNOT gates for the pre-amplification operator rapidly increase from $n=7$ to $n=8$ because the state preparation circuits require $2^n-2$ CNOT gates for implementation. 

\subsubsection{Success probability}
\label{sec:results_1stHamil_prob}
\begin{figure}[ht]
        \centering
        \includegraphics[width=1.0 \textwidth]{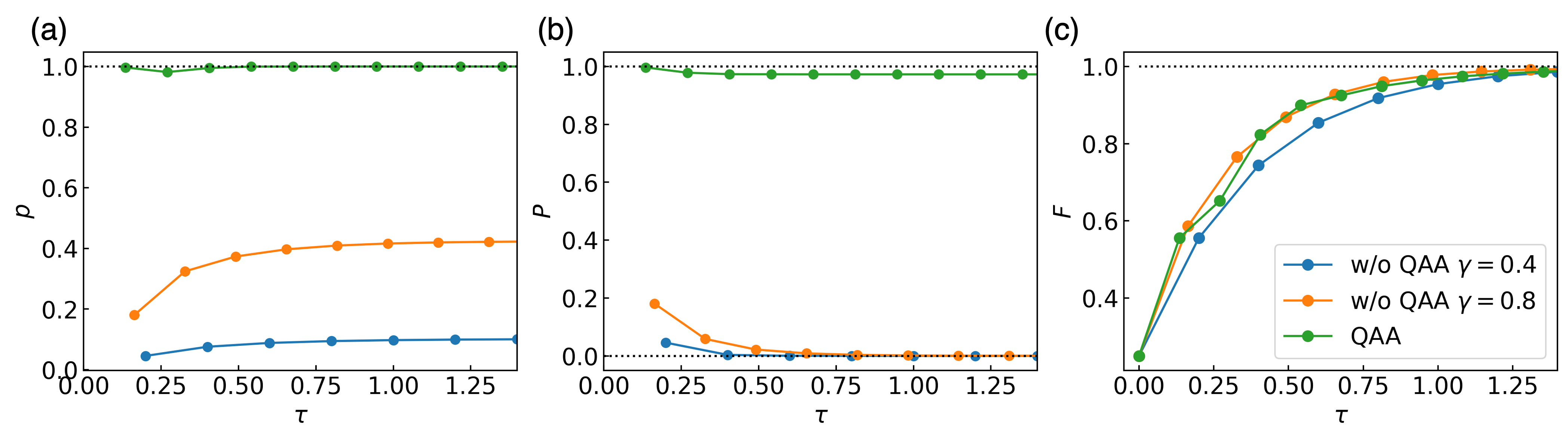}
 \caption{
    The simulation results of PITE with QAA for one-electron confined in the one-dimensional harmonic potential for success probability at each PITE step $p_k$ (a), total probability defined by $P_k=\prod_{\ell=1}^{k} p_{\ell} $ (b), and fidelity $F$ (c) as a function of imaginary-time $\tau$. Blue and orange lines show the results solely performed by PITE with $\gamma=0.4, 0.8$, respectively. Green lines represent the results by PITE combined with QAA. Each imaginary-time step are drawn with points.
 }
 \label{fig:pite_result_fqh}
\end{figure}
We used six qubits to discretize the one-dimensional system to calculate the success probability and fidelity along the imaginary time. 
In the situation, we present the success probability for each PITE step, total success probability, and fidelity depending on the imaginary time in Fig. \ref{fig:pite_result_fqh}. 
As mentioned in the previous subsection, we adopted the different $\Delta\tau$ depending on $\gamma$. 
However, we set the values smaller than $\Delta\tau=0.2$ due to the necessity for the Suzuki-Trotter decomposition in the real-time evolution. 
Specifically, we used $\Delta\tau=0.20, 0.16$, and $0.14$ for solely used PITE with $\gamma=0.4$ and $0.8$, and PITE combined with QAA, respectively. 
We observed the same improvement by QAA for success probability and total success probability in the previous section. In addition, the calculated fidelity is shown in Fig. \ref{fig:pite_result_fqh}(c). 
We measured the number of CNOT gates and circuit depth for the circuits until six steps. 
The number of CNOT gates increases with the number of steps as $(2678,6968,19838,58448,174278,521768)$. 
The circuit depth is as $(4016,10603,30347,89578,267254,800281)$.

In this simulation, the initial state is prepared as the superposition of the ground state and the three lowest excited states, which are obtained from the exact diagonalization of the given Hamiltonian.
The initial state preparation requirs exponential number of CNOT gates. 
Thus, efficient state preparation technique should be explored from the viewpoint of practical applications.

\section{Conclusions}
\label{sec:conclusions}
In this paper, we have reported an algorithm in which the success probability of the probabilistic imaginary-time evolution (PITE) method is improved by combining it with quantum amplitude amplification (QAA). 
In the actual combination of PITE and QAA, we found that by introducing a pre-amplification operator instead of an amplification operator, we can reduce the circuit depth. 
In addition, we found that by optimizing the parameters contained in the PITE operator we can reduce the repetition number of the pre-amplification operators and we can change the probabilistic character of PITE to a deterministic one. 
We successfully demonstrated that the combination of PITE and QAA works efficiently and reported a case in which the quantum acceleration is achieved.
We have found that when the deterministic imaginary-time evolution (deterministic ITE) method is applied within the first few steps, where the success probability is lower than in the later steps, it gives great performance. However, when the deterministic ITE is used for longer ITE steps, the circuit depth becomes deepened exponentially (the drawback of the deterministic ITE).
The construction of the methods for generating a good initial state close to the true solution that may finish the calculations within a few steps should be explored.

\section*{Acknowledgments}
This work was supported by MEXT under "Program for Promoting Researches on the Supercomputer Fugaku" (JPMXP1020200205) and by JSPS KAKENHI under Grant-in-Aid for Scientific Research (A) No. 21H04553. The computation in this paper was done using supercomputer Fugaku provided by the RIKEN Center for Computational Science–Supercomputer Center at the Institute for Solid State Physics at the University of Tokyo.

\appendix

\section{Physical meanings of the failure state in PITE}
\label{estimation_of_error}
In this subsection, we discuss the case where the amplified state by QAA is not a pure separable state but a somehow entangled state containing a small portion of a failure state with a small value, $\varepsilon$, such as:
\begin{gather}
    |\Psi_{\mathrm{in}} \rangle
    \equiv
    \sqrt{\frac{1-\varepsilon}{a}} \mathcal{M} |\psi \rangle \otimes |0\rangle
    +
    \sqrt{\frac{\varepsilon}{1-a}} \sqrt{1-\mathcal{M}^{2}} |\psi \rangle \otimes |1\rangle,
\label{Eq:QITE_with_QAA_2step:psi_w_error}
\end{gather}
where $\langle \psi|\mathcal{M}^{2}|\psi\rangle = a$,  $\langle\Psi_{\mathrm{in}}|I_{2^n}\otimes P_1|\Psi_{\mathrm{in}}\rangle = \varepsilon$, and $P_{j}$ is a projection operator onto the ancillary $|j\rangle$ state.
Let us consider the physical meanings of the failure state. We denote the operators and parameters of PITE in the second step by $\mathcal{M}^{\prime}, \Theta^{\prime}$, and $\kappa^{\prime}$ since the parameters in the second step are different from those in the first step in general.
The entangled state undergoes the PITE operator in the second step as
\begin{gather}
    |\Psi_{\mathrm{in}}\rangle
    \longmapsto 
    \Big(
        \sqrt{\frac{1-\varepsilon}{a}}
        \mathcal{M}^{\prime}
        \mathcal{M}
        +
        \sqrt{\frac{\varepsilon}{1-a}}
        \sqrt{1-\mathcal{M}^{\prime 2}}
        \sqrt{1-\mathcal{M}^2}
    \Big)
    |\psi\rangle \otimes |0\rangle
    \nonumber \\
    +
    \Big(
        \sqrt{\frac{1-\varepsilon}{a}}
        \sqrt{1-\mathcal{M}^{\prime 2}}
        \mathcal{M}
        -
        \sqrt{\frac{\varepsilon}{1-a}}
        \mathcal{M}^{\prime}
        \sqrt{1-\mathcal{M}^2}
    \Big)
    |\psi\rangle \otimes |1\rangle .
\end{gather}
When the measurement outcome of the ancillary qubit is $|0\rangle$ state,
we obtain the success state as
\begin{gather}
    \frac{1}{\sqrt{\mathbb{P}_{0}}}
    \Big(
        \sqrt{\frac{1-\varepsilon}{a}}
        \mathcal{M^{\prime}}
        \mathcal{M}|\psi\rangle
        +
        \sqrt{\frac{\varepsilon}{1-a}}
        \sqrt{1-\mathcal{M}^{\prime 2}}
        \sqrt{1-\mathcal{M}^2}|\psi\rangle
    \Big) ,
\end{gather}
with the probability 
\begin{gather}
    \mathbb{P}_0 
    =
    \langle \psi |
    \Big(
        \sqrt{\frac{1-\varepsilon}{a}}
        \mathcal{M^{\prime}}
        \mathcal{M}
        +
        \sqrt{\frac{\varepsilon}{1-a}}
        \sqrt{1-\mathcal{M}^{\prime 2}}
        \sqrt{1-\mathcal{M}^2}
    \Big)^{2}
    | \psi \rangle .
\end{gather}
If the amplified state by QAA contains a small portion of a failure state, a state acted on by the operator $\sqrt{1-\mathcal{M}^2}$ twice is contaminated in the success state in the next step by a portion of the failure state. 
As shown in Eq. (\ref{PITEwithQAA:output_state_1st_order}), the operator $\sqrt{1-\mathcal{M}^2} $ is approximated in the approximate PITE circuit $\mathcal{C}_{\mathrm{PITE}}^{(1)}$ as
\begin{gather}
	\sqrt{1-\mathcal{M}^2} 
    =
    \sqrt{1-\gamma^{2}}
    \left(
        1 + \mathcal{H} s^{2} \Delta \tau
    \right)
    +
    \mathcal{O}(\Delta\tau^{2})
    =
    \sqrt{1-\gamma^2} e^{\mathcal{H} s^{2} \Delta\tau} 
    +
    \mathcal{O}(\Delta\tau^{2}),
\end{gather}
where we rewrite the operator to the exponential form in the last equality to clearly see the effect of the operation. 
We see the operator $\sqrt{1-\mathcal{M}^2}$ works as an imaginary-time evolution for backward direction in the first order of $\Delta\tau$.

\section{Imaginary-time steps size determination}
\label{appendix:other_parameter_determination}
In this subsection, we discuss how to determine the imaginary-time step $\Delta\tau$. 
In determining $\Delta\tau$, it is necessary to consider the periodicity of the real-time evolution circuit. 
Let $\{\lambda_i\}$ be eigenvalues, the controlled time evolution gate behaves as $e^{-i\lambda_i s\Delta\tau}$ to the input state.
It is desirable to set the rotation angle of the $R_z$ gate smaller than $\pi/4$ for implementing the real-time evolution since the $R_z$ gate is periodic every $\pi/4$ ignoring the global phase. 
For the component showing the largest phase change under the controlled time evolution gate $e^{-i\lambda_{i}s\Delta \tau}$, the condition is expressed as
\begin{gather}
    s\Delta\tau \lambda_{\max}  \leq \frac{\pi}{4} ,
\label{eq:condition_for_dtau}
\end{gather}
where $\lambda_{\max} \equiv \max_{i}|\lambda_{i}|$ is the maximum eigenvalue.
Since it is generally impossible to know $\lambda_{\max}$ in advance, it is preferable to estimate it by an approximation algorithm such as a mean-field approximation.
Let us focus on the max-cut problem, for example, there exists a classical algorithm that finds the solution of an unweighted 3-regular graph with an approximation rate of 93.26\% \cite{Goemans1995, Halperin2004}.
For the numerical simulation in this paper, we assume the maximum eigenvalue as $\lambda_{\max} = 1.2\lambda_{\mathrm{exact}}$, and determine the imaginary-time step  as $\Delta\tau = \pi/(4s\lambda_{\max})$ according to Eq. (\ref{eq:condition_for_dtau}), where $\lambda_{\mathrm{exact}}$ is the exact maximum eigenvalue. 

Besides, the larger $\Delta\tau$ can be permitted for the smaller $\gamma$ because the variable $s$ is an increasing function of $\gamma$ in the range $0<\gamma<1$, as shown in Eq. (\ref{PITE:definition_of_s}).
Of course, if $\Delta\tau$ is too large, the approximation of the first-order approximate PITE circuit will not work, so it should have an upper bound.

Let us summarize the practical procedure for determining the optimal $\Delta \tau$ below. First we determine the $\lambda_{\max}$. Note that the parameters $\{\Delta \tau, \gamma, s\}$ should be self consistent. We prepare the tentative parameters $\{\Delta \tau, \gamma, s\}$ and then measure the $\alpha^2$. From the measured parameter $\alpha^2$, we update the parameters $\{\gamma, s\}$ and subsequently update parameter $\Delta \tau$. Again we measure the $\alpha^2$. We repeat this procedure until the parameter $\Delta \tau$ converges.

\bibliographystyle{quantum}
\bibliography{ref}

\end{document}